\documentclass[review]{elsarticle}

\usepackage{lineno,hyperref}
\usepackage{amsmath, amssymb, mathtools}
\usepackage[caption=false]{subfig}
\usepackage{caption}
\usepackage{appendix}
\usepackage{tabularx}
\usepackage{booktabs}
\modulolinenumbers[5]

\journal{arXiv}

\begin{document}

\begin{frontmatter}

\title{Model selection for identifying power-law scaling}

\author[1,2]{Robert Ton}
\author[1]{Andreas Daffertshofer\corref{mycorrefauthor}}
\address[1]{MOVE Research Institute, Department of Human Movement Sciences, VU Amsterdam, The Netherlands}
\address[2]{Center for Brain and Cognition, Computational Neuroscience Group, Universitat Pompeu Fabra, Roc Boronat 138, Barcelona, 08018, Spain}


\cortext[mycorrefauthor]{Corresponding author}
%

\begin{abstract}
Long-range temporal and spatial correlations have been reported in a remarkable number of studies. In particular power-law scaling in neural activity raised considerable interest. We here provide a straightforward algorithm not only to quantify power-law scaling but to test it against alternatives using (Bayesian) model comparison. Our algorithm builds on the well-established detrended fluctuation analysis (DFA). After removing trends of a signal, we determine its mean squared fluctuations in consecutive intervals. In contrast to DFA we use the values per interval to approximate the distribution of these mean squared fluctuations. This allows for estimating the corresponding log-likelihood as a function of interval size without presuming the fluctuations to be normally distributed, as is the case in conventional DFA. We demonstrate the validity and robustness of our algorithm using a variety of simulated signals, ranging from scale-free fluctuations with known Hurst exponents, via more conventional dynamical systems resembling exponentially correlated fluctuations, to a toy model of neural mass activity. We also illustrate its use for encephalographic signals. We further discuss confounding factors like the finite signal size. Our model comparison provides a proper means to identify power-law scaling including the range over which it is present.
\end{abstract}

\begin{keyword}
\texttt{Model selection, power law, DFA}
\end{keyword}

\end{frontmatter}

\linenumbers
\section*{Introduction}
\noindent
Power laws are a hallmark of systems exhibiting critical dynamics. If a signal's correlation structure is scale-free, i.e. if it does not depend on the temporal or spatial scale of observation, then it displays a power-law structure. Power laws are ubiquitous in nature. They have been observed in  physics \cite{Hurst1951} and biology \cite{Pengetal1995} as well as in economy \cite{Rogers1997,Willingeretal1999,Borland2005} and sociology \cite{Shangetal2008}.  Recent electrophysiological recordings revealed the presence of power laws in nervous activity \cite{LinkenkaerHansenetal2001,Palvaetal2013,Botcharovaetal2015,Tonetal2015}, indicating complex (neuronal) networks operating in a critical state \cite{Chialvo2010, Shaoetal2012}. Neuronal networks with such critical dynamics are believed to have optimal characteristics for neural functioning \cite{Shew&Plenz2013}. 

Scale-free behavior can be identified as a linear relationship in a log-log representation of a signal's power spectral density | hence the phrase 1/f-process | or of its auto-correlation function.\footnote{The auto-correlation function is typically replaced by the mean squared displacement to avoid spurious effects of non-stationarities. This motivated the introduction of the detrended fluctuation analysis explained later on.} If the signal consists of successive increments, the linear slope $\alpha$ in the log-log representation may be identified as the Hurst exponent $H$, with $H<0.5$ and $H>0.5$ marking negative and positive correlations, respectively. For $H=0.5$ the (integrated) signal resembles Brownian motion (i.e. a Wiener process), i.e. a random walk whose increments stem from a (uncorrelated) Gaussian white noise process. Particularly interesting is the case of $H>0.5$, in which the auto-correlation function decays slower than the exponential auto-correlation function of Brownian motion. In this case of so-called persistent behavior, the extent of correlation (or 'memory') is increased, which is the reason why these processes are also being referred to as containing long-range correlations. 

The most common algorithmic implementation to determine power-law behavior is  detrended fluctuation analysis (DFA) \cite{Pengetal1994}. In DFA one estimates the signal's fluctuation magnitudes $F(n)$ as a function of interval size $n$ after removing linear (or non-linear) trends per interval. In the presence of a power law, the linear slope of $F$ as a function of $n$ in a log-log representation is the corresponding scaling exponent $\alpha$. 

While DFA has proven robust when it comes to confounding (weakly non-linear) trends \cite{Huetal2001} or non-stationarities \cite{Dang&Molnar1999}, it does not provide any means to determine whether a power law is present or not. Deviations from power-law behavior, however, are common and may originate from different dynamical mechanisms. Since the slope $\alpha$ is typically estimated via simple regression, the corresponding coefficient of determination, $R^2$, may serve to quantify the goodness-of-fit of linearity. However, this measure is quite insensitive \cite{Anscombe1973,Grech&Mazur2013} and above all it does not allow for readily specifying the range in which the power-law behavior is likely to be present. That range might not only be limited by the finite size of observation. When a signal is contaminated by a non-trivial, non-linear trend (e.g., a sinusoid), the fluctuation plots may have a scale-dependent slope \cite{Huetal2001}. Then, the slope of the linear regression over a (too) large range does not necessarily represent the scaling behavior of interest \cite{Bardet&Kammoun2008,Chenetal2002,Botcharovaetal2013}. In view of the variety of such confounders, many applications still rely on mere visual inspection to determine violations of the linearity assumption. 

We here present an assessment of power-law behavior based on proper model comparison. Rather than averaging the mean squared fluctuations over consecutive intervals, we approximate their density function to estimate the maximum likelihood function underlying the common model selection benchmarks, namely Akaike or Bayesian information criteria. The aim is to identify the presence of power-law behavior, its corresponding scaling exponent and the range over which it is valid.

\section{Methods}
\noindent
Starting with a discrete time series $X(t)$ with $t = 1, \ldots, N$, we compute its cumulative sum $Y(t)=\sum_{\tau=1}^tX(\tau)$, which is considered the signal\footnote{If the process generating $X(t)$ falls in the category of fractional Gaussian noise, the cumulative sum will represent fractional Brownian motion.}. In DFA one divides $Y(t)$ into non-overlapping intervals of size $n$ yielding ${\lfloor{N/n}\rfloor}$ signals $Y_i \left( t \right)$ with $i = 1, \ldots, \lfloor{N/n}\rfloor;~t = 1, \ldots, n$. The notation $\lfloor \cdot \rfloor$ refers to the floor function. Since the focus is on the analysis of the fluctuation structure of the signal, one first removes the signal's trend $Y^{\text{trend}}_i \left( t \right)$ over the interval $i$ before quantifying the mean squared fluctuations $F_i \left( n \right)$ in every interval as\footnote{In the current study we only consider linear trends in line with the original form of DFA \cite{Pengetal1994} but we note that this approach may be readily generalized to non-linear trends \cite{Bundeetal2000}.} 
\begin{align}
F_i \left( n \right) = \sqrt{ \frac{1}{n} \sum_{t = 1}^{n} \left[ Y_i \left( t \right) - Y^{\text{trend}}_i \left( t \right) \right]^2 } \ . \label{Fi}
\end{align}
\noindent%
These fluctuation magnitudes are further averaged over consecutive intervals
\begin{align*}
\bar{F} \left( n \right) = \sqrt{ \frac{1}{{\lfloor{N/n}\rfloor}} \sum_{i=1}^{\lfloor{N/n}\rfloor} F_i^2 \left( n \right) } \ .
\end{align*}
A power law in the signal's autocorrelation structure is present if the averaged mean squared fluctuation structure is independent of the scale at which it is observed. Let $b$ be an arbitrary base, then scale-freeness implies that rescaling $b$ by $n$ changes $\bar{F}$ only by some factor, i.e. $\bar{F}(n\cdot{b}) = n^\alpha\bar{F}(b)$, with $\alpha$ the scaling parameter. In a log-log representation this simplifies to $\log\bar{F}(n\cdot{b}) = \alpha\log{n}+\log\bar{F}(b)$. In DFA one hence seeks to detect power laws by identifying a straight line in the log-log plot of the averaged mean squared fluctuations as a function of interval size:
\begin{align}
\log \bar{F}\left(n\right) =  \alpha  \log{n} \label{LLPL} + \log\bar{F}_0 \ .
\end{align}
The scaling parameter $\alpha$ agrees with the Hurst-exponent $H$ when considering signals like fractional Gaussian noise. The scaling exponent $\alpha$ is typically identified as the slope of a linear fit determined using linear regression.
 
 \subsection{Approach}
 \noindent%
Albeit implicitly, by the mere use of linear regression when identifying scale-free correlations, one already assumes the presence of a power law. We here advocate to test this assumption statistically. We will employ a model selection approach using conventional information criteria to compare the linear model against alternatives. The most commonly used information criteria are the Akaike (AIC) and Bayesian (BIC) information criteria \cite{Burnham&Anderson2002}, which are defined as
\newcounter{eq_dummy}\setcounter{eq_dummy}{1}%
\renewcommand{\theequation}{\arabic{equation}\alph{eq_dummy}}%
\begin{align}
\mbox{AIC}_c &= -2 \ln  \mathcal{L}_{\text{max}}  + 2K + \frac{2K(K+1)}{M-K-1} \label{AIC} \\ 
\addtocounter{equation}{-1}\addtocounter{eq_dummy}{1}
\mbox{BIC} &= -2 \ln  \mathcal{L}_{\text{max}} + K\ln{M}\ ; \label{BIC} 
\end{align}
\renewcommand{\theequation}{\arabic{equation}}%
$K$ represents the number of parameters of the model under study, $M$ the number of intervals with different size $n$, and $ \mathcal{L}_{\text{max}}$ denotes the maximum value of the likelihood function $\mathcal{L}$, which quantifies the goodness-of-fit. 
AIC$_c$ and BIC can be regarded as asymptotic approximations to the log model evidence \cite{Penny2012}. The log model evidence may be decomposed into accuracy (the first terms above) and model complexity. The latter basically scores the number of free parameters used to provide an accurate explanation for the data. Maximizing model evidence (minimizing AIC$_c$ or BIC) therefore provides an accurate and minimally complex explanation for data.
That is, when applied to a set of candidate models, the model with the least information criterion value is the one that establishes this optimal compromise between goodness-of-fit and parsimony.\footnote{For the sake of legibility we restrict the main text to the report of BIC | the very comparable AIC$_c$ results can be found in {\it Appendix A}. We used the finite sample size correction AIC$_c$ of the conventional AIC.}

\begin{figure}[!ht]
\centering 
\subfloat[]{\label{subfig:fig1a}\includegraphics[width=0.45\textwidth]{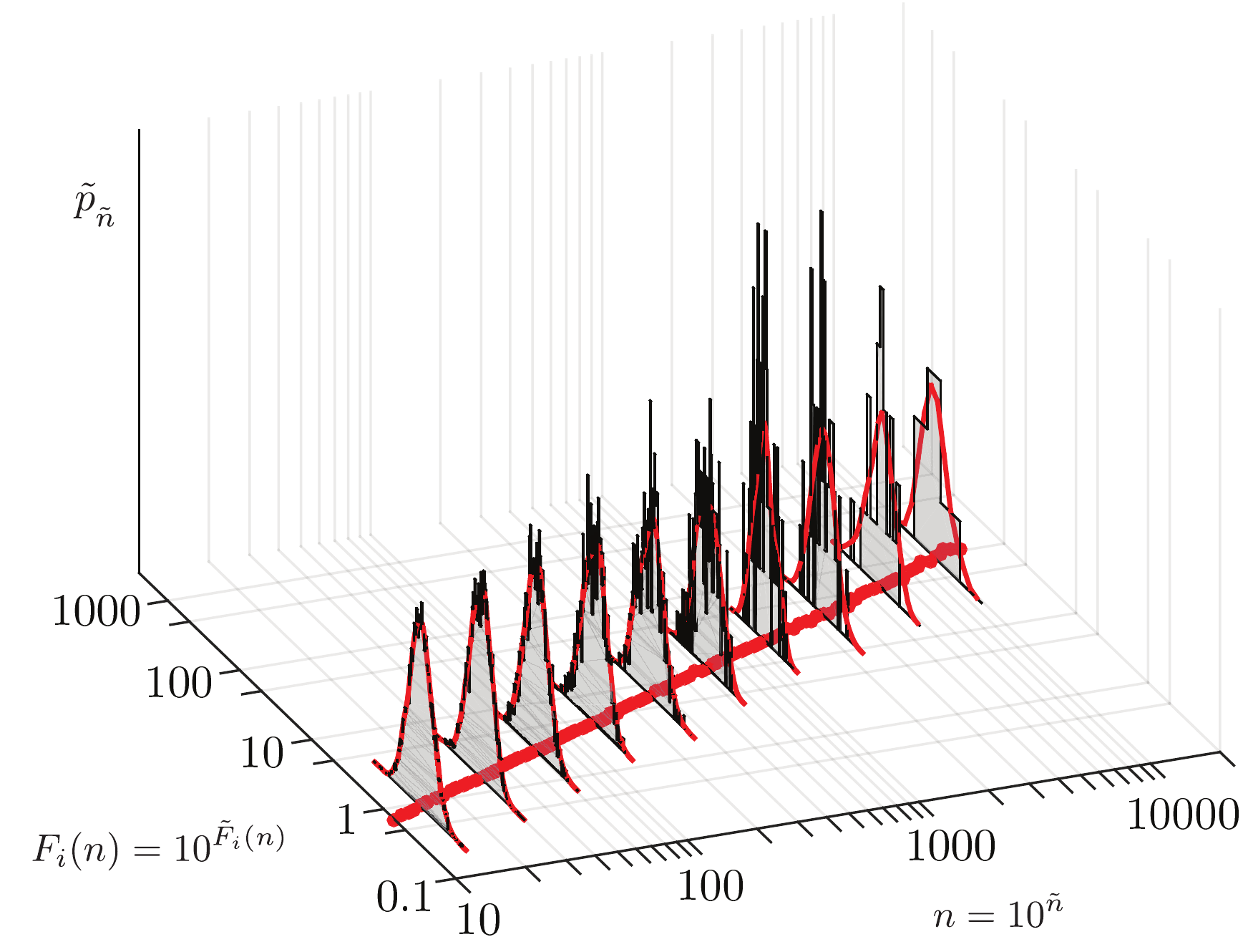}}
\subfloat[]{\label{subfig:fig1b}\includegraphics[width=0.45\textwidth]{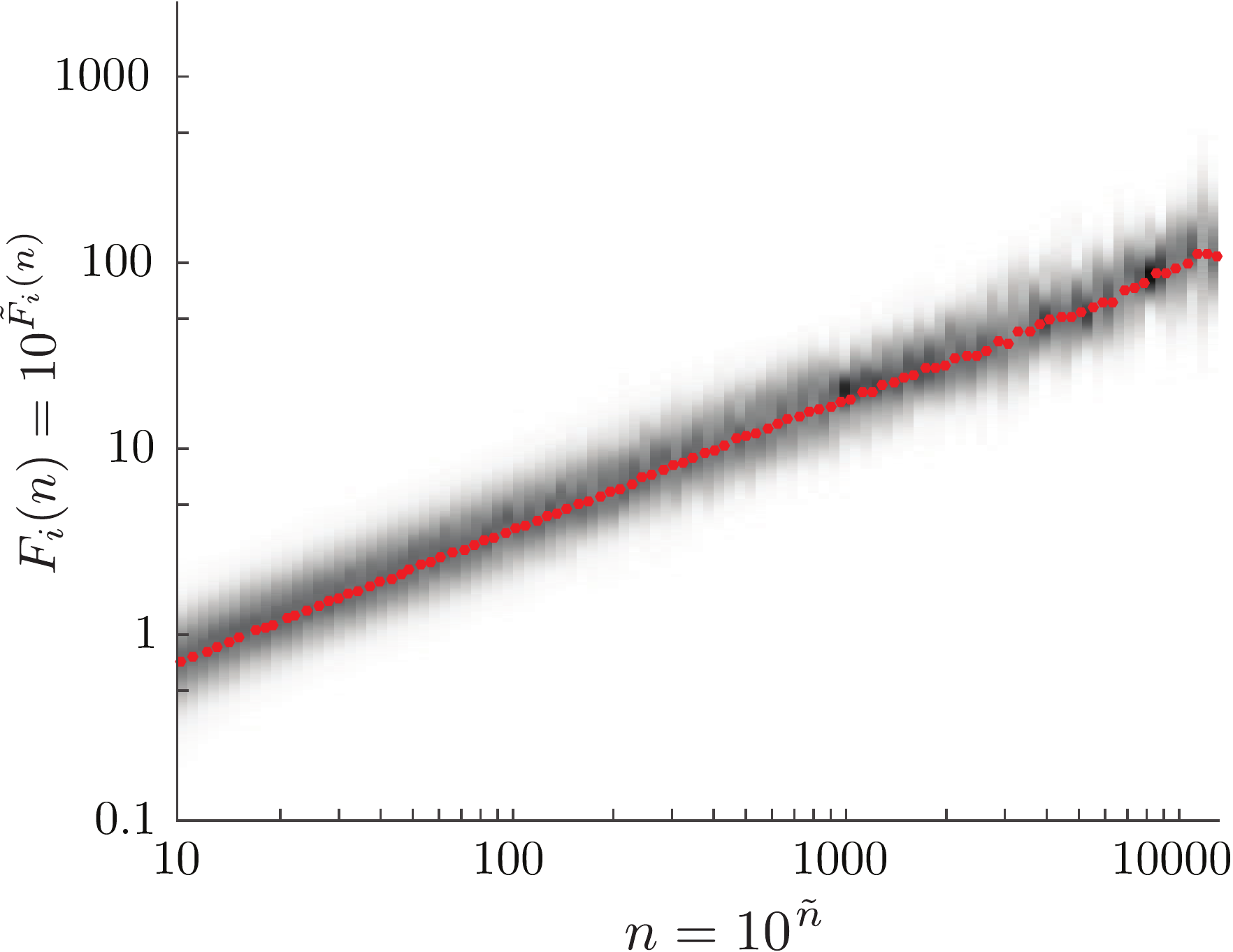}}
\caption{\small {\bf Fig. \ref{subfig:fig1a}:} Results for a fractional Gaussian noise processes with $H\!=\!0.7$ to sketch procedures. The figure shows 
the probability densities $p_n$ of $F_i$ for different interval sizes $n$; black outer lines are the histograms, the red lines are the corresponding kernel density estimates. The red dots at the bottom of the plot represent the expectation values. {\bf Fig. \ref{subfig:fig1b}}: Display of the same results from the top ($n$ along the horizontal and estimates $F_i(n)$ on the vertical axis. Red dots depict the expectation values $\left\langle F_i \right\rangle$ on which a regression might be based as reminiscent to conventional DFA. Gray areas represent the densities $p_{n}$ with darker colors indicating higher values. Note that in the following all log-likelihood estimates are based on these types of densities. We note that throughout the paper the figures display $\tilde{F}_i$ values but we relabelled axes to correspond to their non log-transformed counterparts. 
}
\label{fig:fig1}
\end{figure}

Model comparison requires a proper estimate of $\mathcal{L}_{\text{max}}$. For this we interpret $F_i$ as a 'stochastic' variable and determine the corresponding probability density $p_{n}(F_i)$. We do this via a kernel density estimation procedure using the set of $\left\lfloor{N/n}\right\rfloor$ realizations $F_i$ given by \eqref{Fi}. 
This non-parametric approach has the advantage that it does not prescribe any form of $p_{n}$ and allows $p_{n}$ to acquire different forms depending on interval size $n$. Next, since all subsequent fitting will be based on log-log  coordinates, we introduce the log-transformed variables $\tilde{n} = \log{n}$ and $\tilde{F}_i  = \log F_i$. 
Just as $p_{n}(F_i)$ being the probability density corresponding to $F_i$, we find that $\tilde{F}_i$ is distributed according to $\tilde{p}_{n}( \tilde{F}_i)$.
We illustrate this in Fig. \ref{fig:fig1}, where we display the densities $\tilde{p}_{n}$ along with the corresponding histograms of $\tilde{F}_i$. 
As an equivalent of the averaged $\bar{F}_i$ values, we there also report the expectation values with respect to $\tilde{p}_{n}$:
\begin{align}
\mathbb{E}\left[\tilde{F}_i\right] = \left \langle \tilde{F}_i \right \rangle = \int_{\mathbb{R}^+} x \cdot \tilde{p}_{n}(x)\, dx \label{EF}
\end{align}
In case of a non-symmetric density, the expectation value will be different from the value with highest probability; on the right-hand side of \eqref{EF} the variable $x\in\mathbb{R}^+$  covers the state space of $\tilde{F}_i$.

In the model selection we aim at finding a model $f_{\theta} \left( \tilde{n} \right)$ that properly describes $\tilde{F}_i$. The candidate models are functions $f_{\theta}$ parametrized by the set $\theta$. 
For example, these could be linear (power law) functions of scale or more elaborate (polynomial) functions (see below).
The likelihood function $\mathcal{L}$ is defined as the product of the afore-defined probability densities, evaluated at the model values $f_{\theta} \left(\tilde{n}\right)$: 
\begin{align}
\ln \left( \mathcal{L}\Big(\theta | \tilde{F}_i \right) \Big) = \ln \left( \prod_{n} \tilde{p}_{n}\left(f_\theta \right) \right) =  \sum_{n} \ln \Big( \tilde{p}_{n} \left( f_\theta \right)\Big) \ . \label{L}
\end{align}
The density $\tilde{p}_{n}$ is evaluated at the values given by model $f_{\theta}$ and thus quantifies the probability that the model value $f_{\theta}$ is contained in the set of realizations $\tilde{F}_i$ for a given interval size $n$. 
Since the information criteria require maximized log-likelihood values, we subsequently maximize $\ln \left( \mathcal{L} \right)$ by determining the set $\theta_{\text{max}}$, i.e. $\ln \left( \mathcal{L}_{\text{max}} \right) =  \ln \left( \mathcal{L}\left(\theta_{\text{max}} | \tilde{F}_i \right) \right)$. Note that the set $\theta_{\text{max}}$ is specific for the candidate model $f_{\theta}$ under consideration.
The parameters $\theta_{\text{max}}$ correspond to the maximized likelihood and can be interpreted as the most probable set of parameters corresponding to a particular model given the distributions $\tilde{p}_n$ of fluctuation magnitudes.

\newpage
\subsection{Implementation}
\noindent
We implemented the entire procedure in Matlab (version 2015a, The Mathworks, Natwick, MA). All the source code including a working example is available at \url{www.upmove.org}.

We used different types of simulated signals to evaluate the performance of our method. In all cases we generated 1000 realizations of length $N=2^{17}$, unless stated otherwise. We also added beamformed magneto-encephalographic (MEG) signals recorded during rest \cite{Cabraletal2014} to show the applicability of our approach to neurophysiological data. Here we only analyzed the envelope dynamics at a single (virtual) channel and refer to \cite{Tonetal2015} for more detail.

\subsubsection{Fractional Gaussian noise} 
\noindent
Seminal generators for self-similar signals are fractional Gaussian noise (fGn) processes, introduced by Mandelbrot \& Van Ness  \cite{Mandelbrot&VanNess1968}. A fGn process can be realized through increments of fractional Brownian motion that is given by
\begin{align}
\begin{split}
B_H(t) - B_H(0) &= \int_{-\infty}^t K(t-s) \,dB(s)\\
\text{with}\qquad
K(t-s) &= {\scriptsize\begin{cases} (t-s)^{H-\frac{1}{2}} & \text{for} \ \ 0 \leq s \leq t \\ (t-s)^{H\!-\!\frac{1}{2}} - (-s)^{H\!-\!\frac{1}{2}} & \text{for} \ \ s < 0 \end{cases}}\quad\text{and}\quad 0\!<\!H\!<\!1
\end{split} \label{fBm}
\end{align}
where $dB(s)$ indicates a stochastic integral with respect to conventional Gaussian white noise. The increments of $B_H(t)$ are equivalent to the aforementioned fGn process and we denote them as $\Delta B_H(t)$. The case $H=\frac{1}{2}$ represents non-correlated increments, whereas  $H<\frac{1}{2}$, $H>\frac{1}{2}$ indicate processes with anti-correlated and correlated increments, i.e. anti-persistent or persistent behavior, respectively. Realizations of $\Delta B_H(t)$ for $0\!<\!H\!<\!\frac{1}{2}$ were generated using \cite{Lowen1999} and for ($\frac{1}{2}\!<\!H\!<\!1$) via a truncated symmetric moving average filter. 
We simulated signals with Hurst exponents $H = \{$ 0.1, 0.3, 0.5, 0.7, 0.9 $\}$. 

\subsubsection{Potential function | introducing deterministic features}
\begin{figure}[h!]
\centering 
\includegraphics[width=0.75\textwidth]{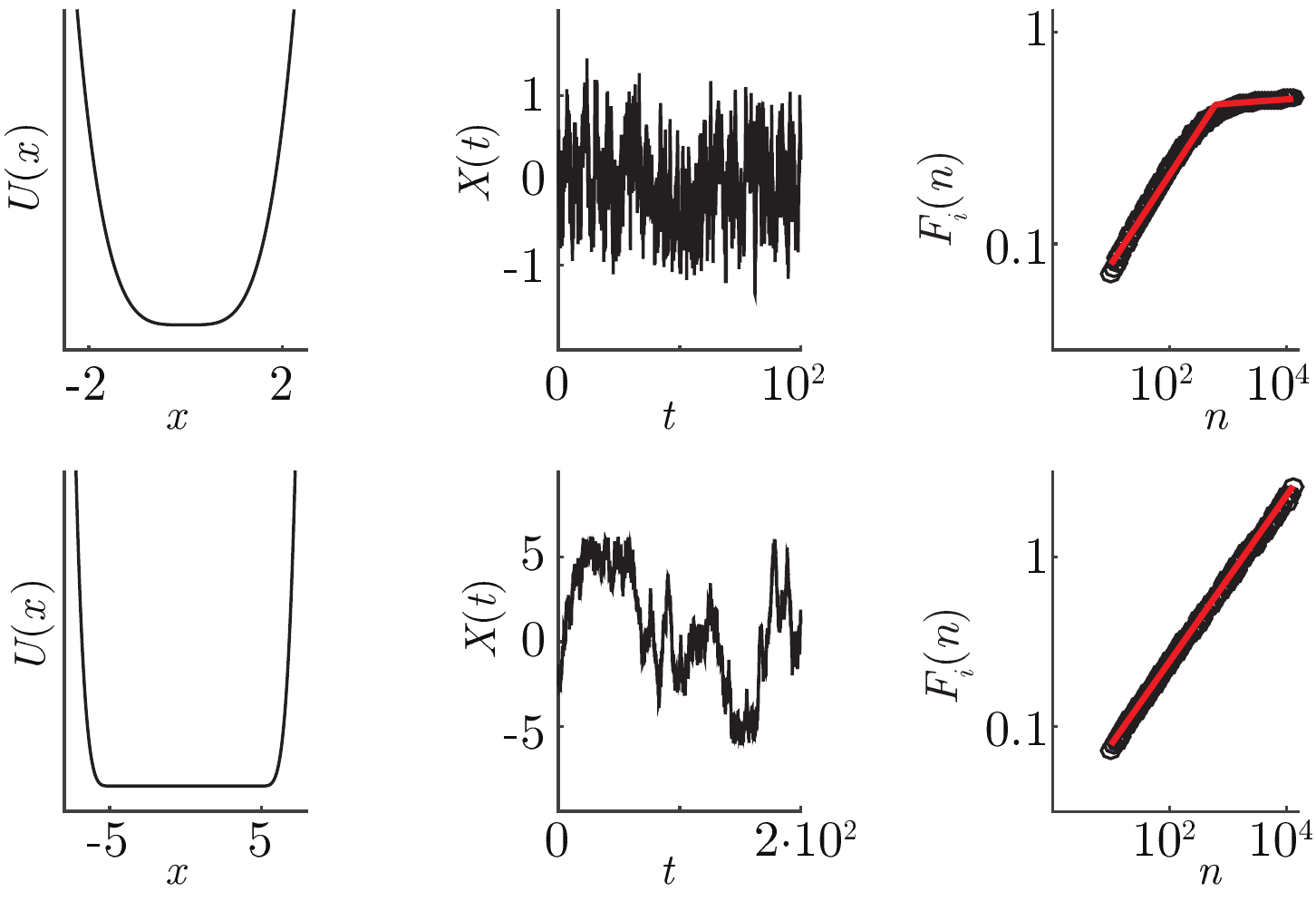}
\caption{\small{\bf Fig. \ref{fig:fig2}}: Two sample paths $X(t)$ and their fluctuation plots are depicted in Fig. \ref{fig:fig2} for $w = 0$ and $w = 5$ to illustrate the limitation of power-law scaling due to the presence of an attractive potential force. In the upper row $w = 0$ yields a flattening of the $\tilde{F}_i$ curve for larger interval sizes. There the potential is narrow, i.e. the local attraction is strong. When the potential well is wide enough, as is the case for $w = 5$, this flattening occurs outside the here-considered range of interval sizes. Then, the log-log fluctuation plot still appears linear for this scaling range (lower right panel). 
The Hurst exponent for the noise process was $H = 0.5$, i.e. white Gaussian noise.}
\label{fig:fig2}
\end{figure}

\noindent
To test how our approach could handle deviations from power-law scaling we used signals $\Delta B_H(t)$ in the presence of a saturating dynamics. The generating stochastic differential equation reads
\begin{align}
\begin{split}
dX(t) &= -U' dt + dB_H(t) \\
\text{with}
& \qquad U(x) = {\tiny\begin{cases} 0 & \text{for} \ \ \lvert x \rvert \leq w \\ \left \lvert x - w \right \rvert^4 & \text{for}\ \ \lvert x \rvert \geq w \end{cases}}
\end{split} \label{fGnPotSDE}
\end{align}
where $U$ represents the potential function with $w$ a threshold value indicating the width of the potential well; the prime denotes differentiation with respect to $x$. The potential $U$  introduces a deterministic component in the signal bounding the displacement of the sample path $X(t)$ and thereby its fluctuations $F_i$. The system is not influenced by the potential as long as $ \left \lvert X(t) \right \rvert \leq w$. This means that as long as $w$ is large enough compared to the maximum interval size, the bounding effect of $U$ will be invisible. Integration was performed using an Euler-Maruyama scheme with step-size ${\rm{d}}t = 0.01$; see Fig. \ref{fig:fig2}.

\subsubsection{A toy model for neural mass dynamics}
\noindent
We further considered a class of stochastic processes proposed by \cite{Ruseckas&Kaulakys2010}. The signals $X(t)$ were generated by the stochastic differential equation \eqref{Kaulakyseqctu} defined in {\it Appendix A}. Important for the application in neuroscience is that equation \eqref{Kaulakyseqctu} has been derived from a point process model with stochastic inter-pulse intervals \cite{Kaulakys&Ruseckas2004}, which may be interpreted as the firing of a single neuron.
This is illustrated in Fig. \ref{subfig:fig6a} for a single realization $X(t)$ whose erratic behavior bears resemblance with neuronal firing. However, a single realization of this process can certainly not be considered relevant when it comes to the description of neural masses, because they represent activity of large populations of neurons. Neural mass activity is considered to underly M/EEG signals that | as mentioned in the Introduction | do display power-law behavior. Therefore we also analyzed averages over multiple realizations $X(t)$. All time series $X(t)$ for this class of signals consisted of $N = 10^6$ samples. 

\subsubsection{Envelope dynamics of beamformed MEG}
\noindent
We used MEG signals that were sampled at 1 kHz during about five minutes resting state (eyes closed) in ten subjects. After down-sampling to 250 Hz, signals were beamformed onto a ninety node brain parcellation \cite{Tzourioetal2002}. In line with previous work, e.g., \cite{LinkenkaerHansenetal2001}, we considered the alpha frequency band (8-12 Hz) of a single occipital source. Details about the data acquisition and pre-processing can be found in \cite{Cabraletal2014,Tonetal2015}; see also {\it Appendix B}.

\subsubsection{Density estimation} 
\noindent
From the different signals, the values $F_i$ were computed according to \eqref{Fi} and their corresponding densities were estimated by a kernel density estimation approach using normal kernels 
\cite{Azzalini&Bowman1997}. 
The number of kernels was adjusted to the number of available values $F_i$ according to $\mbox{min}(100,{\lfloor{N/n}\rfloor})$. 
The $F_i$ were computed for a vector of logarithmically spaced interval sizes in the range $n \in$ [10, N/10] with $M=99$ different interval sizes. 

\subsubsection{Candidate models}
\noindent
Before specifying a set of candidate models we would like to note that any finite selection comes with arbitrariness, at least to some degree. When defining a set of models, however, it is generally recommended to keep the set concise \cite{Burnham&Anderson2004}. 
Given our search for a linear relationship between $\tilde{F}_i$ and $\tilde{n}$, i.e.
\begin{align}
\log F_i =  \alpha  \log{n} + \log{ F_{i,0} } \ \ \Rightarrow \ \  
\tilde{F}_i =  \alpha \,\tilde{n}  + \tilde{F}_{i,0} \ .
\end{align}
we used as a first candidate the linear model:
\setcounter{eq_dummy}{1}\renewcommand{\theequation}{\arabic{equation}\alph{eq_dummy}}%
\begin{align}
f_\theta^1(x) &= \theta_1 + \theta_2 x \ .
 \label{models1}
\end{align}
 \addtocounter{equation}{-1}\addtocounter{eq_dummy}{1}
As alternatives we also tested all possible polynomials up to third order, i.e.
\begin{align}
f_\theta^2(x) &= \theta_1 + \theta_2 x^2 \\  \addtocounter{equation}{-1}\addtocounter{eq_dummy}{1}%
f_\theta^3(x) &= \theta_1 + \theta_2 x + \theta_3 x^2 \\  \addtocounter{equation}{-1}\addtocounter{eq_dummy}{1}%
f_\theta^4(x) &= \theta_1 + \theta_2 x^3 \\ \addtocounter{equation}{-1}\addtocounter{eq_dummy}{1}%
f_\theta^5(x) &= \theta_1 + \theta_2 x + \theta_3 x^3 \\ \addtocounter{equation}{-1}\addtocounter{eq_dummy}{1}%
f_\theta^6(x) &= \theta_1 + \theta_2 x^2 + \theta_3 x^3 \\ \addtocounter{equation}{-1}\addtocounter{eq_dummy}{1}%
f_\theta^7(x) &= \theta_1 + \theta_2 x + \theta_3 x^2 + \theta_4 x^3 \ .
\end{align}
\addtocounter{equation}{-1}\addtocounter{eq_dummy}{1}%
The next two models were derived from the expressions of the variance of a process generated by a (un)stable linear stochastic dynamics: 
\begin{align}
f_\theta^8(x) &= \theta_1 + \theta_2 e^{\theta_3 x} \\ \addtocounter{equation}{-1}\addtocounter{eq_dummy}{1}%
f_\theta^9(x) &= \theta_1 + \frac{1}{\ln(10)}  \ln \left( \theta_1 \left( 1 - e^{- \theta_2 e^{\ln(10) x}} \right) \right) \ .
\end{align}
\addtocounter{equation}{-1}\addtocounter{eq_dummy}{1}%
The exact forms of the above expressions, in particular that of $f_{\theta}^9$, result from transforming the variance expressions into the log-log coordinate system. 

Finally, we  considered a piece-wise linear function because this type of model is frequently used to characterize critical behavior in motor control, in particular postural sway \cite{Collins&DeLuca1994, Bouletetal2010}. That model obeys the form
\begin{align}
\begin{split}
f_\theta^{10}(x) &= \begin{cases} \theta_1 +\theta_2 x & x \leq \theta_4 \\ C + \theta_3 x & x > \theta_4 \\  \end{cases}
\end{split} \label{models4} \qquad\text{with}\quad C = \theta_1 + (\theta_2 - \theta_3) \theta_4 \ .
\end{align}
\renewcommand{\theequation}{\arabic{equation}}%

\subsubsection{Model selection} 
\noindent
Since $\ln\mathcal{L}_{\text{max}}$ in \eqref{BIC} represents the maximum log-likelihood, one has to determine the sets $\theta_{\text{max}}$  for every model $f_{\theta}$. The optimization was performed using a Nelder-Mead simplex search algorithm \cite{Lagariasetal1998}. The parameters resulting from the least squares fit based on the averaged values $\bar{F}$ served as initial values, because we considered this to be an appropriate first approximation to the optimal parameters $\theta_{\text{max}}$. Subsequently, we randomly chose five additional initial conditions to test whether the simplex search ended in a local maximum.

\subsection{Effect of the number of interval sizes} 
\noindent
By the definition of $ \mathcal{L} $ in \eqref{L} combined with \eqref{BIC} the model selection results may in general depend on $M$, i.e. on the number of intervals with different sizes $n$. This is because $\ln \mathcal{L} $ scales approximately with $M$, whereas the $K$-dependent term in \eqref{BIC} does not. Hence the relative contribution of the latter to the BIC decreases with increasing $M$. The smaller $M$, the more model assignment will be biased towards underfitting. We investigated such size effects using signals generated via the potential model \eqref{fGnPotSDE} with $w=2.5$ and distinct values of $M$.

\section{Results}
\noindent
As mentioned above we restrict ourselves to presenting the BIC results, the corresponding AIC$_c$ results can be found in {\it Appendix A}.
\begin{figure}[h!]
\centering 
\includegraphics[width=0.75\textwidth]{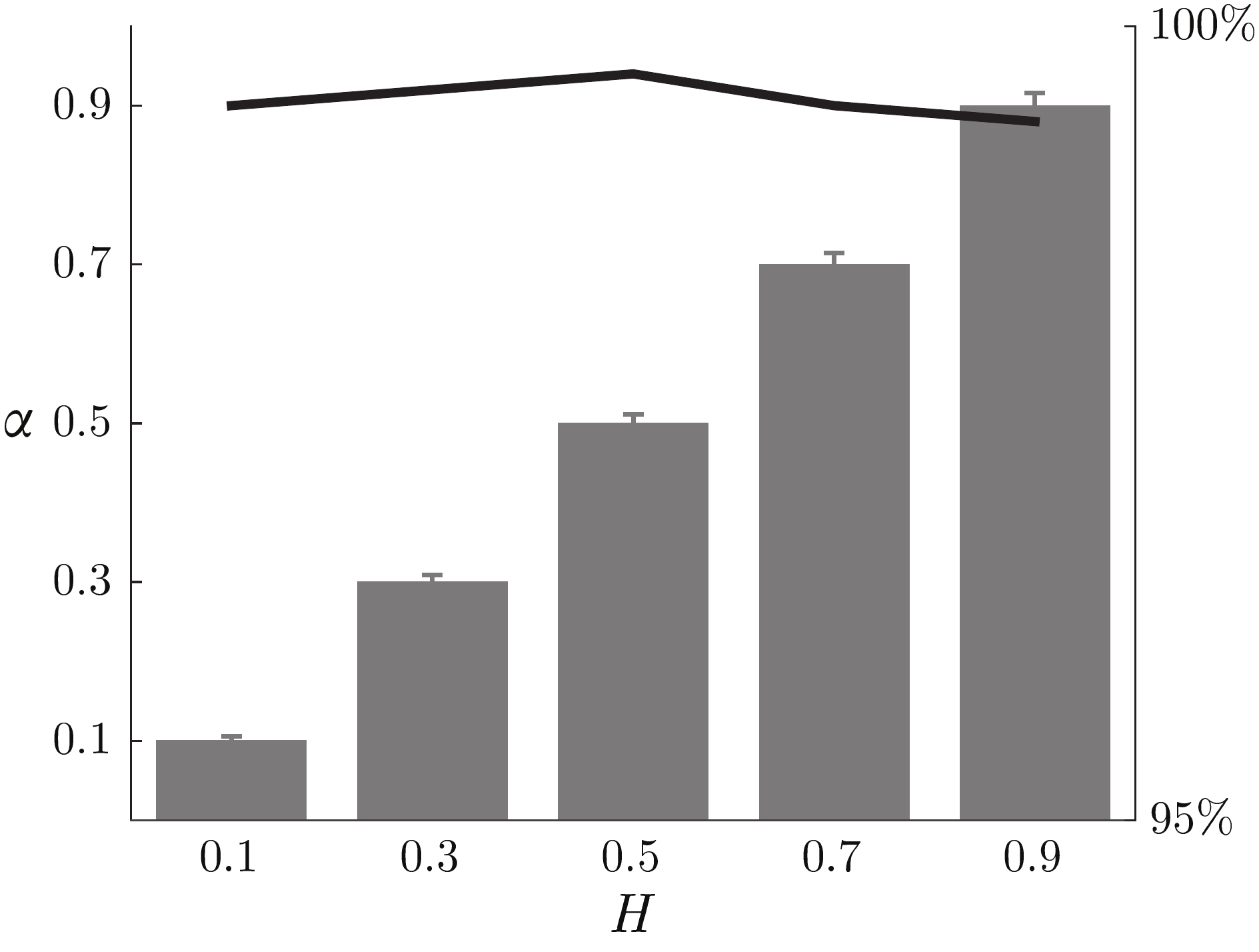}
\caption{\small {\bf Fig. \ref{fig:fig3}}: Performance for fGn processes for the values $H$ depicted on the horizontal axis. Proportion of the cases in which the linear model was preferred is depicted by the line and corresponds to the right vertical axis. On the left vertical axis $\alpha$ denotes the averaged scaling exponent estimate over all realizations that resulted in an assigned linear model. Absolute standard deviations are given as error bars. } 
\label{fig:fig3}
\end{figure}

\subsection{Fractional Gaussian Noise}
\noindent
In Fig. \ref{fig:fig3} we summarize the results for the fGn. More than 99\% of realizations yielded a linear model, i.e. the expected power-law behavior. The estimated scaling exponents $\alpha$ were very close to the simulated Hurst exponents $H$ with relative errors $ \frac{H - \alpha}{H}$ being [1.1\%, 0.3\%, 0.1\%, 0.1\%, 0.1\%]  for $H=$ [0.1, 0.3, 0.5, 0.7, 0.9] respectively. Relative standard deviations of the estimates $\alpha$ amounted to [4.2\%, 2.7\%, 2.2\%, 1.9\%, 1.7\%]. Our approach can thus be considered robust and accurate in identifying and characterizing this class of self-similar signals. The proportions of realizations assigned as power-law as well as the $\alpha$ values closely resemble the results in \cite{Botcharovaetal2013}.

\subsection{Potential function | introducing deterministic features}
\begin{figure}[h!]
\centering
\includegraphics[width = 0.75\textwidth]{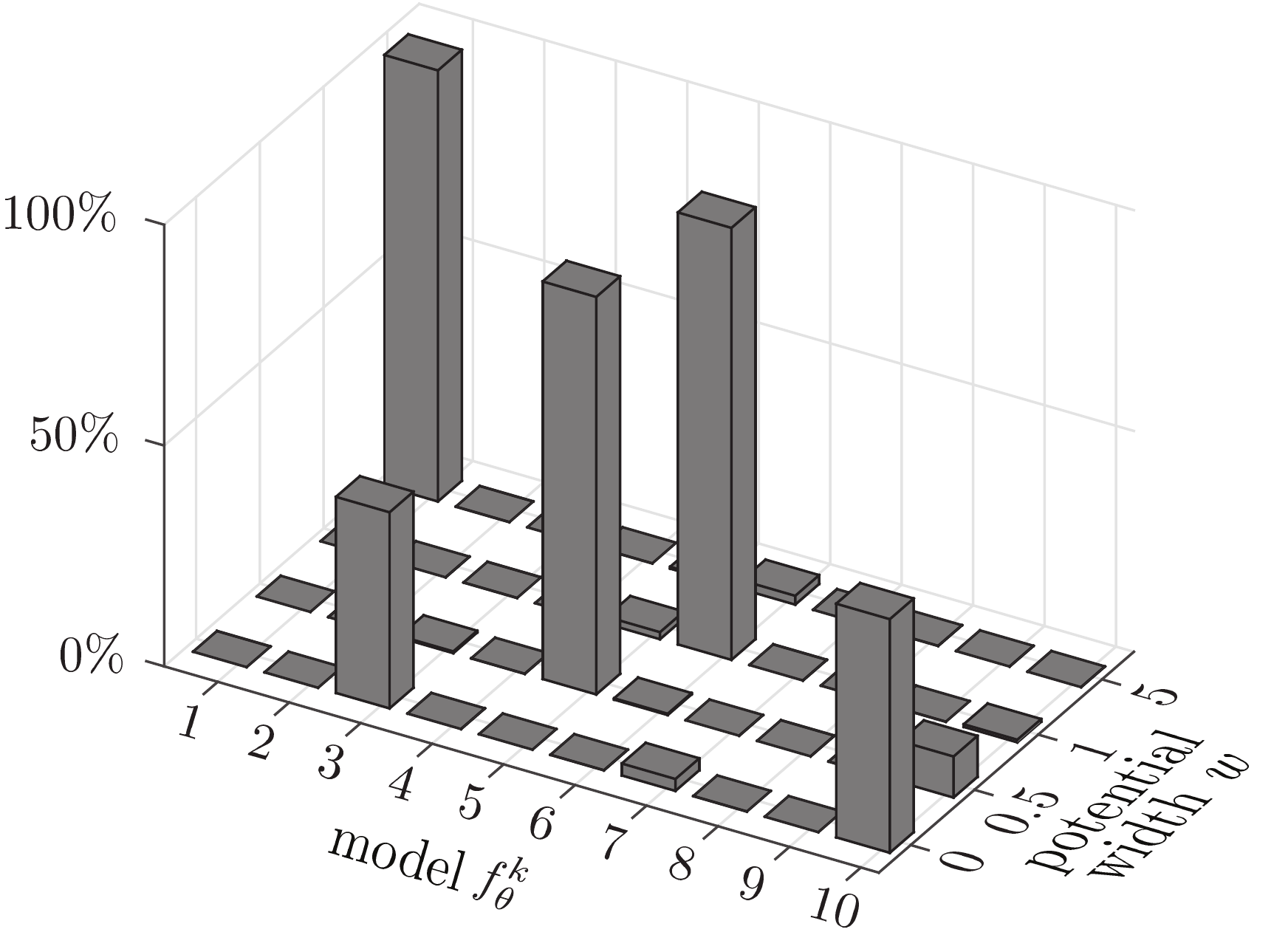} 
\caption{\small {\bf Fig. \ref{fig:fig4}:} Proportions of the simulated realizations for which the indicated model was preferred for four different $w$ values; see Fig. \ref{fig:fig2} for typical sample paths in the case of $w=0$ and $w=5$. Numbers on the non-labeled axis correspond to different forms of the model $f_{\theta}$ defined in \eqref{models1}-\eqref{models4}. The Hurst exponent of the noise process was $H=0.5$. }
\label{fig:fig4} 
\end{figure}

\noindent
Fig. \ref{fig:fig4} summarizes the proportion of realizations assigned to a model of a given form for four different values of $w$. For small $w$ the potential function $U$ bounds the sample paths $X(t)$ and therefore the fluctuations $F_i$ do not scale as a power law. For large $w$ the here-employed range of interval sizes turned out to be insufficient to show this bounding effect. We here discuss the two extreme cases $w = 0$ and $w = 5$. For the first the assigned models were either the piece-wise linear function $f_{\theta}^{10}$ (53\% of realizations) or the second-order polynomial $f_{\theta}^3$ (44\%).  In case of the wide potential $w = 5$, the large majority of realizations was assigned to the linear model (98\%) suggesting the presence of a power law. However, for a larger scaling range, the log-log fluctuation plots that correspond to the $w = 5$ condition would have been classified as non-linear.

\subsection{Effect of number of interval sizes} 
\begin{figure}[h!]
\centering
\subfloat[]{\label{subfig:fig5a}\includegraphics[width=0.32\textwidth]{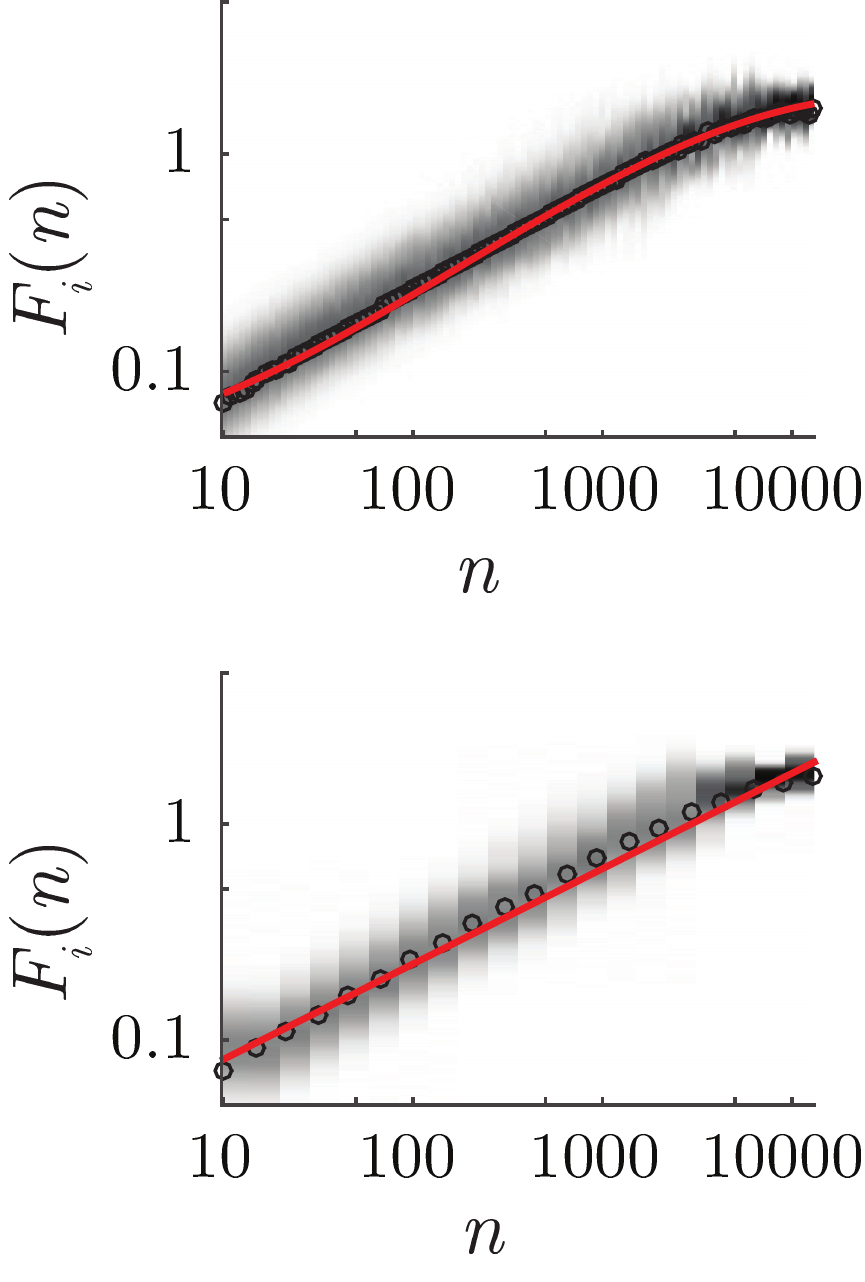}}
\subfloat[]{\label{subfig:fig5b}\includegraphics[width = 0.63\textwidth]{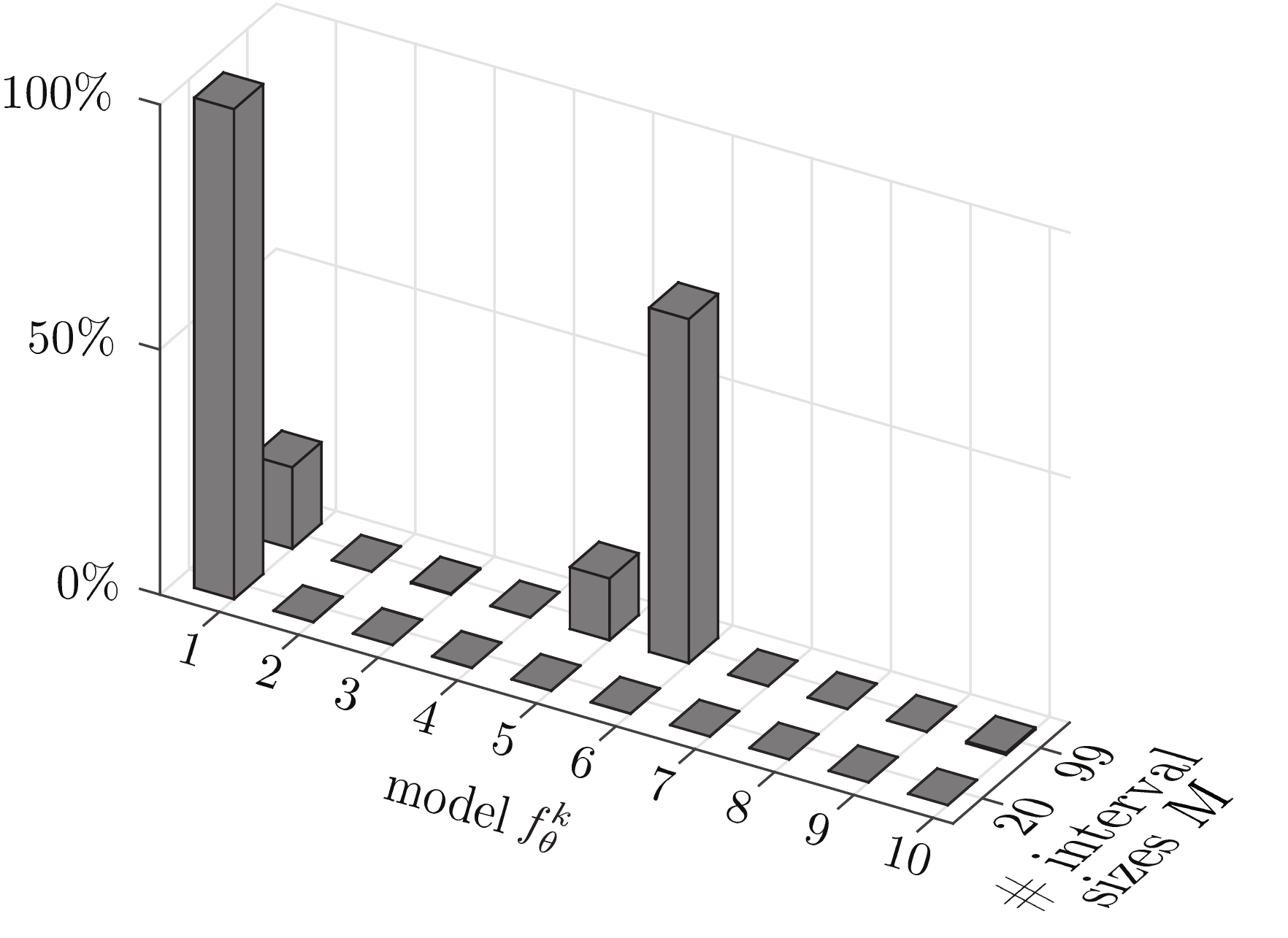}}
\caption{\small {\bf Fig. \ref{subfig:fig5a}:} Typical DFA results of the process given by \eqref{fGnPotSDE} with $w = 2.5$ and $H=\frac{1}{2}$. Colors form a heatmap for the $\tilde{p}_{n}$ values, circles indicate $\left \langle \tilde{F}_i \right \rangle$ values and red lines the optimal fits. For $M = 99$ (upper panel) model $f_{\theta}^6$ was favored, for $M = 20$ the linear function $f_{\theta}^1$. {\bf Fig. \ref{subfig:fig5b}:} Similar as Fig. \ref{fig:fig4} but as function of $M$ for fixed $w = 2.5$. Proportions  of the total number of 1000 realizations for which the indicated model was preferred as function of the number of interval sizes $M$. The same realizations were used for both $M$ values. Numbers on the non-labeled axis correspond to different forms of the model $f_{\theta}$ defined in \eqref{models1}-\eqref{models4}. }
\label{fig:fig5} 
\end{figure}

\noindent
We compared $M=20$ versus our default $M=99$ interval sizes. The results for all realizations are depicted in Fig. \ref{subfig:fig5b}. For $M = 99$, only 16\% of the realizations were assigned to the linear model, but this was markedly different for $M = 20$. There all realizations were assigned to the linear model. Note that this reflects the sole effect of the number of interval sizes $M$ as all included realizations agreed for both $M$-values.

\subsection{Toy model for neural mass dynamics}
\begin{figure}[h!]
\centering
\subfloat[]{\label{subfig:fig6a}\includegraphics[width=0.32\linewidth]{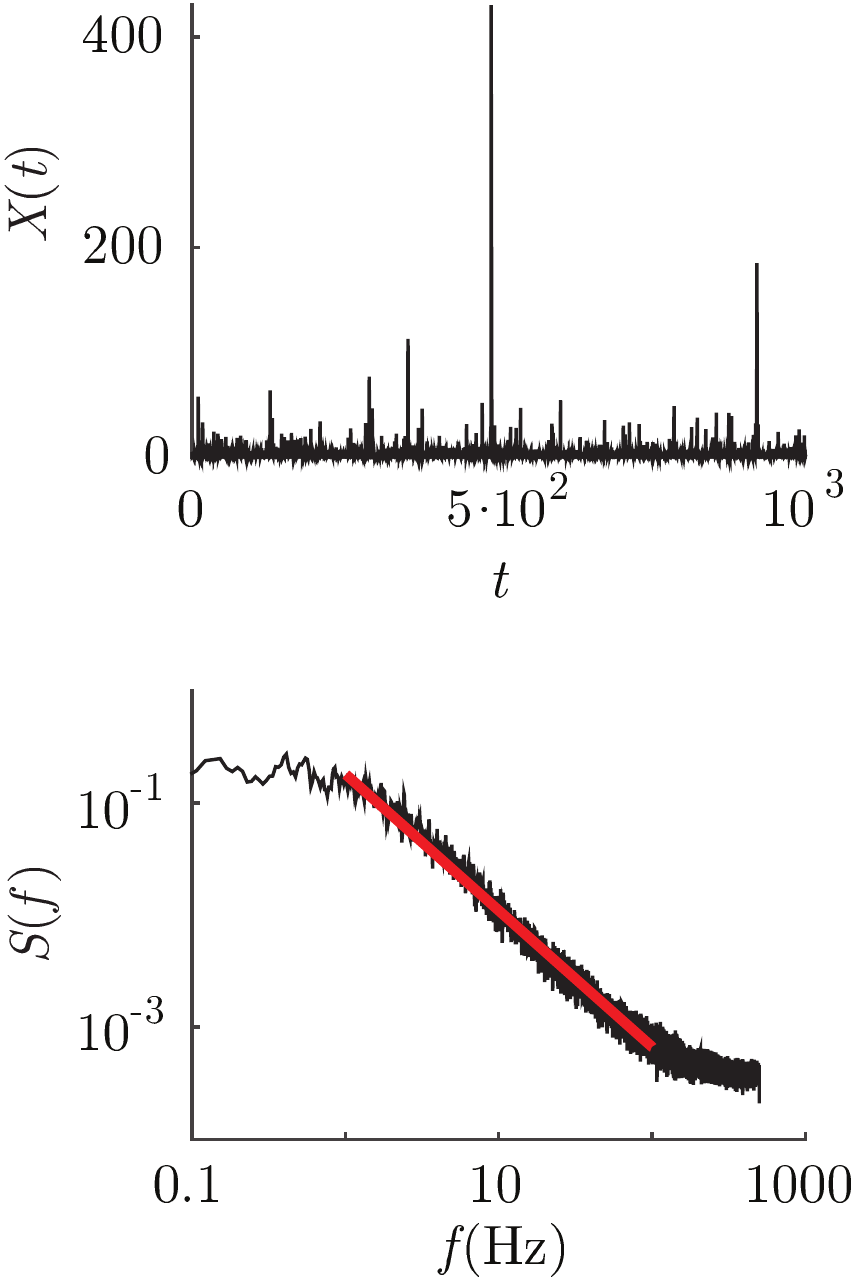}}
\subfloat[]{\label{subfig:fig6b}\includegraphics[width = 0.63\linewidth]{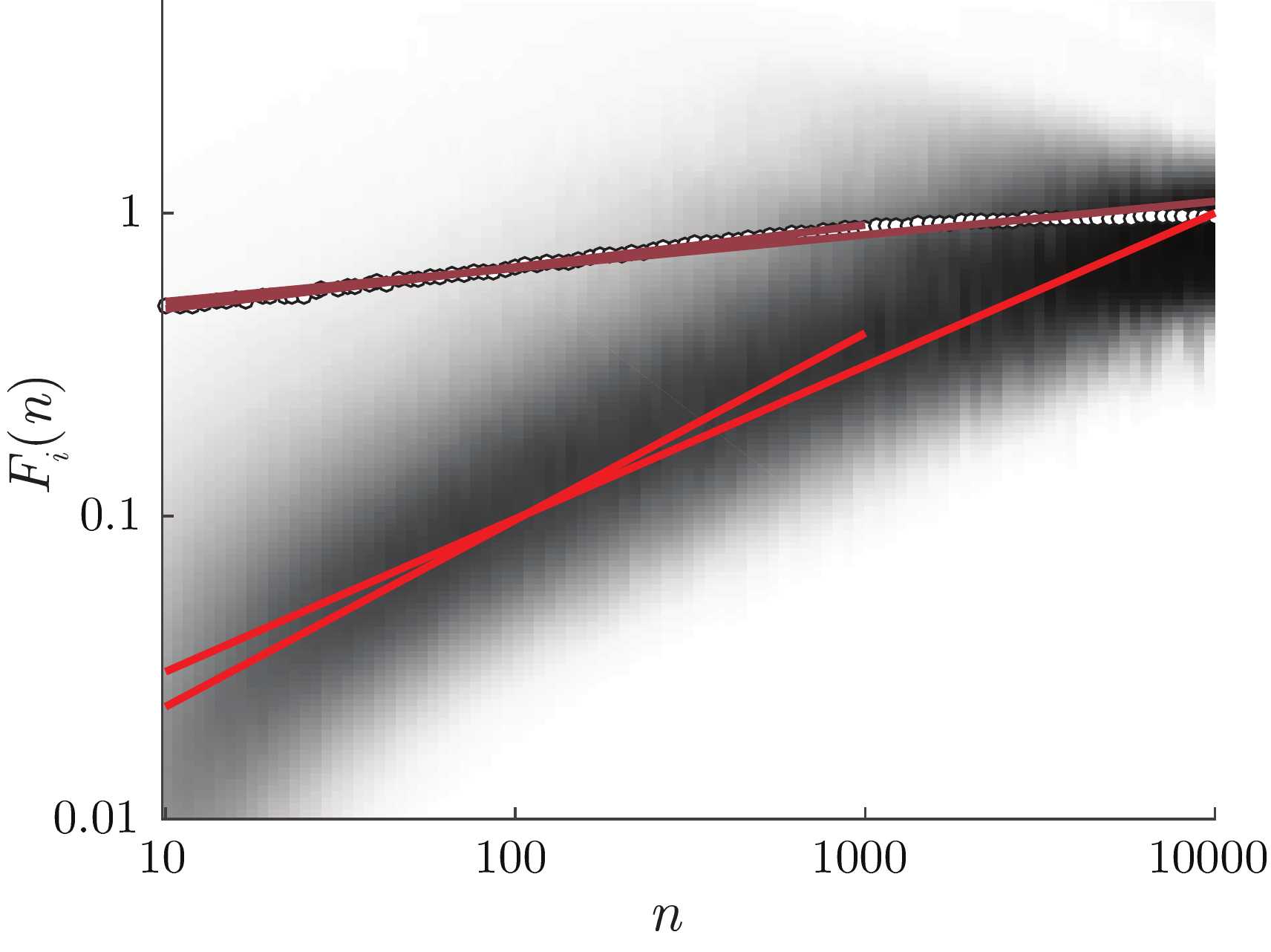}}
\caption{\small {\bf Fig. \ref{subfig:fig6a}:} A single time series $X(t)$ generated by the equation \eqref{Kaulakyseqctu} (upper panel) and its power spectrum (lower panel) with parameter settings $\sigma=1$, $\eta = 2$ and $\lambda = 4$. The power spectrum was determined as a power spectral density estimate using Welch's method with 20 50\% overlapping windows. Assuming a form $S(f) = f^{-\beta}$, the slope of the fit (red line) gives an estimate for $\beta$, which here amounted to $\beta = 1.22$. In case of fractional Brownian motion the exponent $\beta$ is related to the Hurst exponent $H$ by means of $\beta = 1 + 2H$ \cite{Daffertshofer2011}, yielding here $\alpha=0.11$. The fit was determined over the range $f=[1~100]$ Hz corresponding to $n=[10~1000]$. Note that fitting the power spectrum in this manner is equivalent to fitting the averaged $\bar{F}_i$ values. {\bf Fig. \ref{subfig:fig6b}:} The distributions $\tilde{p}_n$ together with the maximum likelihood fit (red line) over two different ranges of window lengths: $n = [10~10^3]$ and $n = [10~10^4]$. White dots represent the averaged values $\bar{F_i}$ with the red-gray lines indicating the fits over the same two window length ranges as above. }
\label{fig:fig6} 
\end{figure}

\noindent
In Fig. \ref{subfig:fig6a} we show a single realization $X(t)$ with the erratic behavior typical for neuronal spike trains. The corresponding fluctuation plot is given in Fig. \ref{subfig:fig6b} together with fits obtained from either conventional linear regression based on the mean values $\bar{F_i}(n)$ and the here proposed fits based on the likelihood function $\mathcal{L}$. The extreme events in Fig. \ref{subfig:fig6a} resulted in a large difference between $\bar{F_i}$ and the modes of the densities $p_{n}$. As a result the exponent estimates differed significantly with $\alpha=0.14$, $\alpha=0.11$ for linear regression versus $\alpha=0.62$, $\alpha=0.57$ for the maximum likelihood estimate with interval lengths ranging over $n = [10~10^3]$ and $n = [10~10^4]$ respectively. Neither of the estimates coincided with the expected value of $H=0.25$ given by \eqref{Hurstrelation} for these parameter settings.

\begin{figure}[h!]
\centering
\subfloat[]{\label{subfig:fig7a}\includegraphics[width=0.32\linewidth]{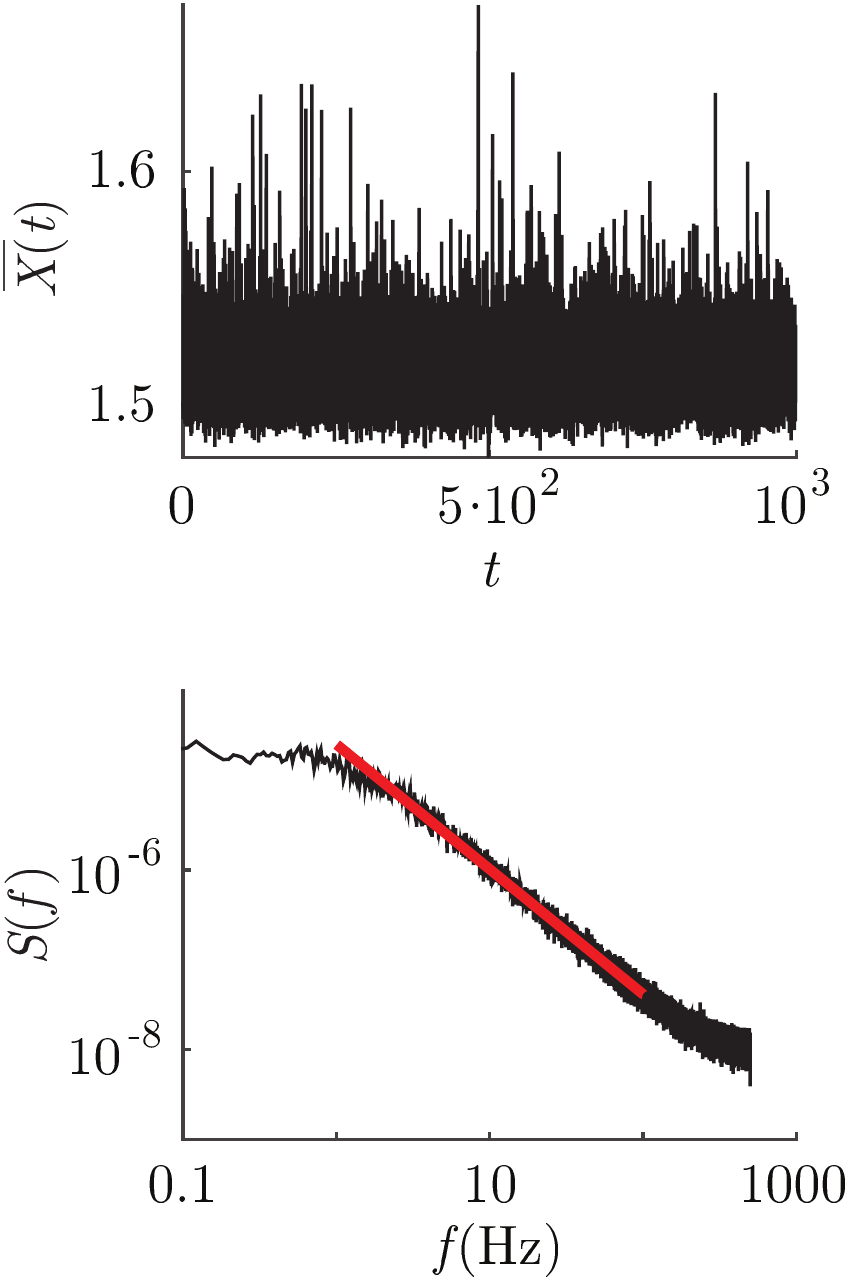}}
\subfloat[]{\label{subfig:fig7b}\includegraphics[width = 0.63\linewidth]{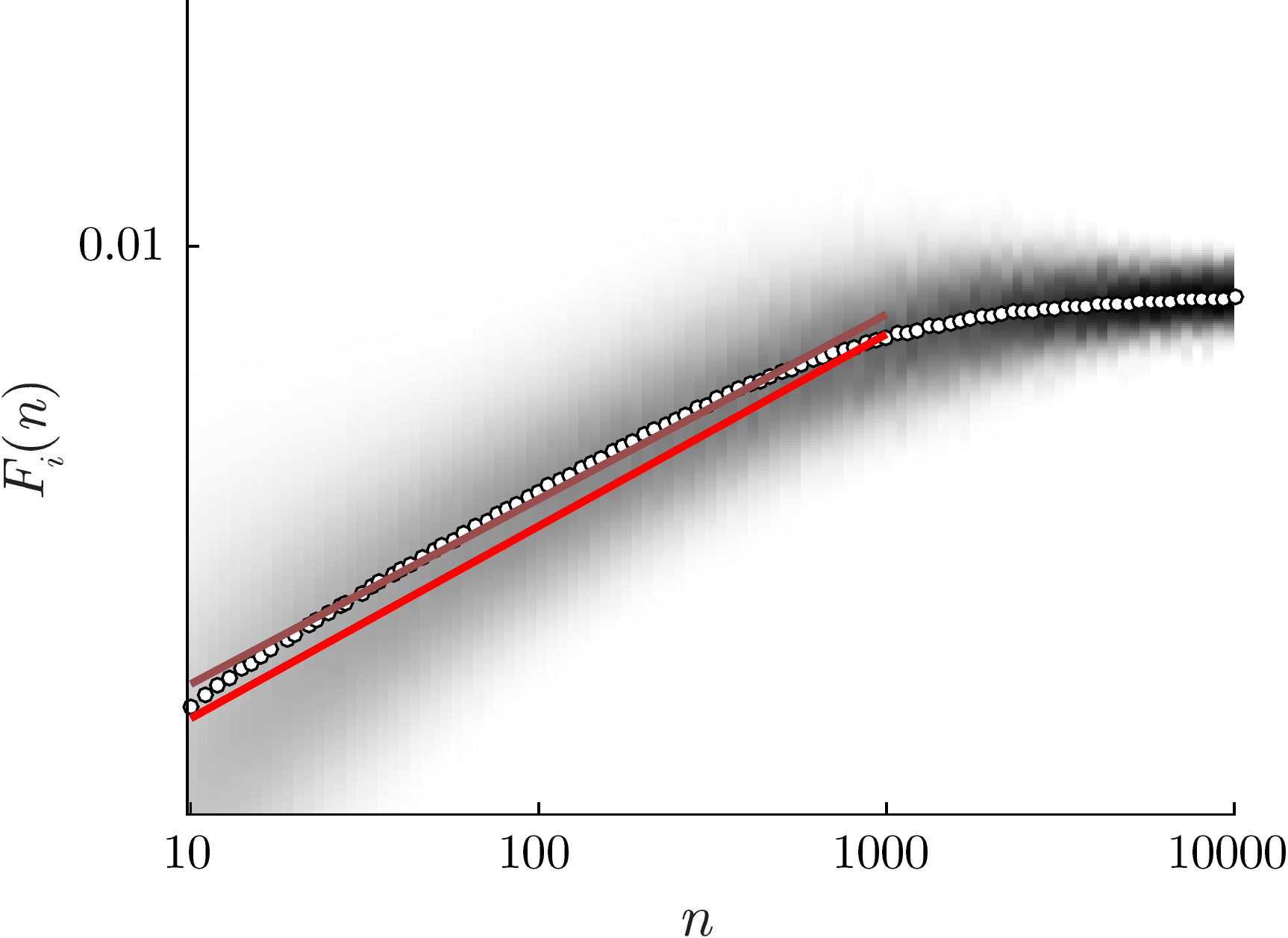}}
\caption{\small {\bf Fig. \ref{subfig:fig7a}:} The averaged signal $\bar{X}(t)$ over 10000 realizations (upper panel) and its power spectrum (lower panel) generated by \eqref{Kaulakyseqctu} with parameter settings $\sigma=1$, $\eta = 2$ and $\lambda = 4$. The $\beta$ exponent estimate was $\beta = 1.39;~\alpha=0.20$, cf. Fig. \ref{subfig:fig6a}. {\bf Fig. \ref{subfig:fig7b}:} The distributions $\tilde{p}_n$ (gray shading)  together with the maximum likelihood fit (red line, $\alpha = 0.24$). White dots represent the averaged values $\bar{F}_i$ with the red-gray lines indicating the fit over the same  window length range as above ($\alpha=0.23$). The obtained scaling exponent estimates were close to the theoretical value $H=0.25$. }
\label{fig:fig7} 
\end{figure}

As said we sought to mimic neural mass activity and simulated 10000 realizations $X(t)$ that we averaged to obtain $\bar{X}(t)$. We display this signal together with its power spectral density in Fig. \ref{subfig:fig7a};  we removed the first 1000 samples were removed to discard transients. Due to averaging, the prominent spikes in $X(t)$ almost vanished in $\bar{X}(t)$. The scaling exponent estimates in the range $n = [10~10^3]$ ($\alpha=0.23$ and $\alpha=0.24$ for linear regression and the MLE respectively) were close to the theoretical value \cite{Ruseckas&Kaulakys2010}. In Fig. \ref{subfig:fig7b} we restricted the analysis to the range $n = [10~10^3]$, because for the larger range of window sizes $n = [10~10^4]$ the log-log fluctuation plot was classified as non-linear.

\begin{figure}[h!]
\centering
\subfloat[]{\label{subfig:fig8a}\includegraphics[width=0.32\linewidth]{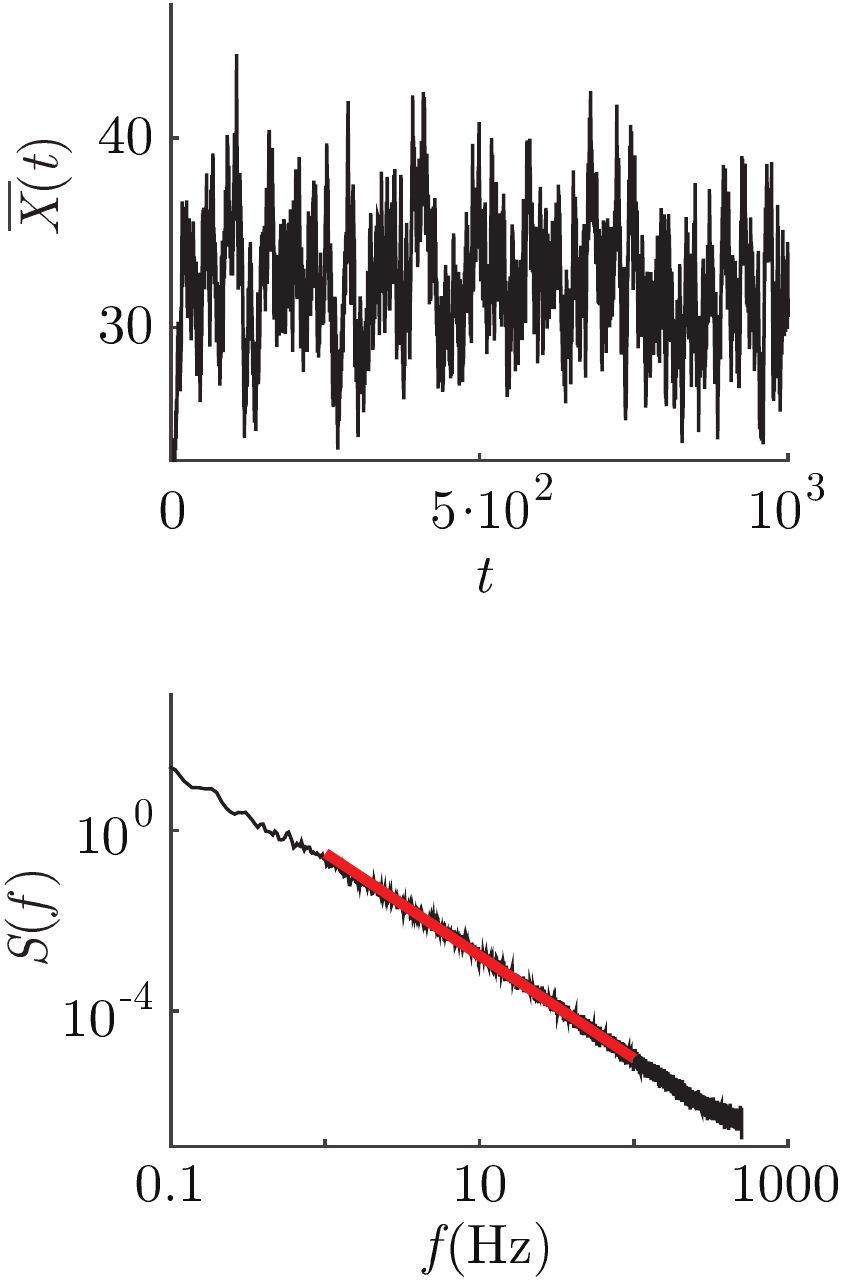}}
\subfloat[]{\label{subfig:fig8b}\includegraphics[width=0.63\linewidth]{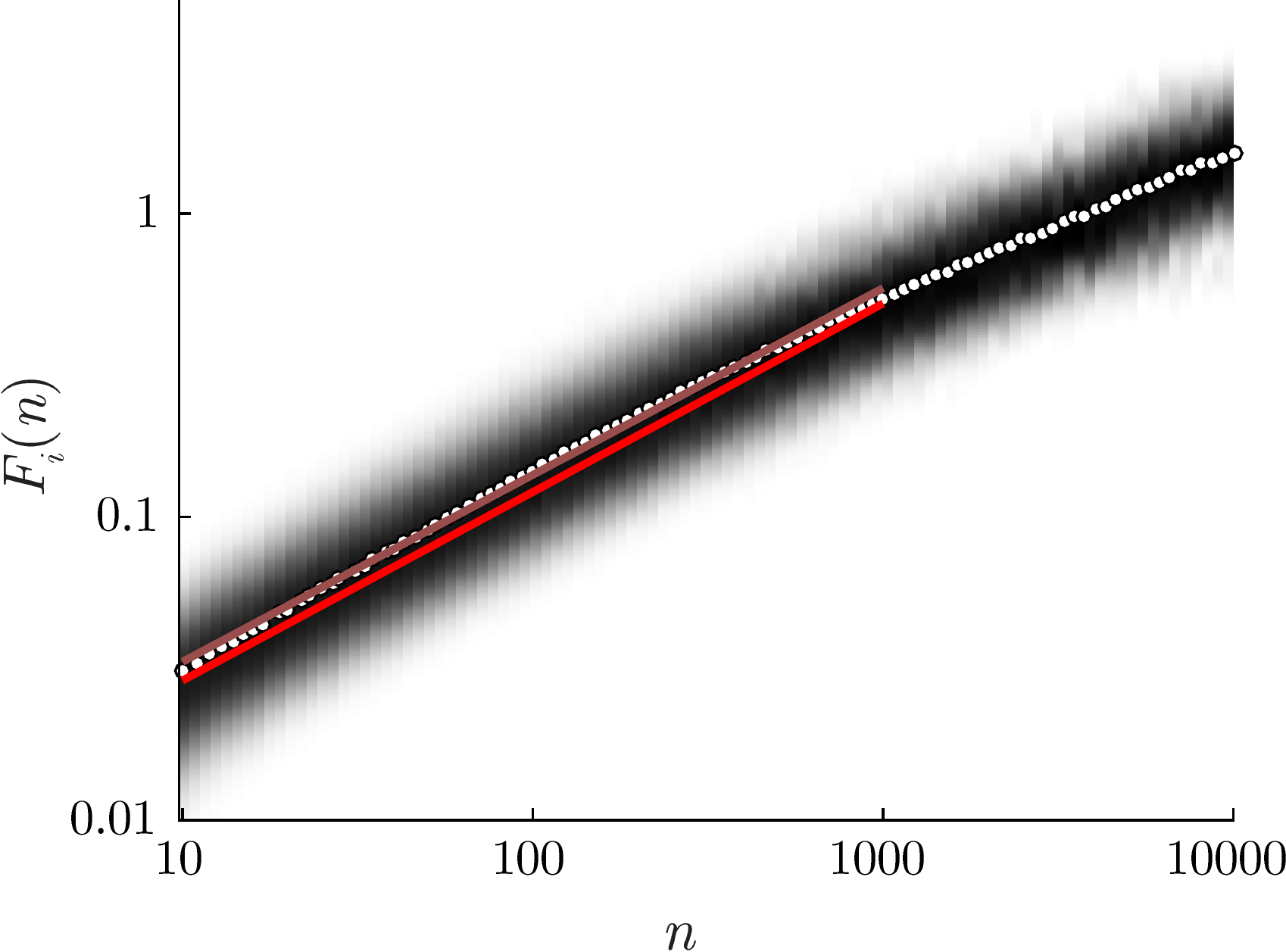}}
\caption{\small {\bf Fig. \ref{subfig:fig8a}:} The average $\bar{X}(t)$ over 1000 realizations $X(t)$ (upper panel) and its power spectrum (lower panel) with parameter settings $\sigma=1$, $\eta = 2$ and $\lambda = 4$ using \eqref{Kaulakyseqctu}. The scaling exponent was $\beta=2.27;~\alpha = 0.63$, cf. Figs. \ref{subfig:fig6a} \& \ref{subfig:fig7a}. {\bf Fig. \ref{subfig:fig8b}:} The distributions $\tilde{p}_n$ together with the maximum likelihood fit (red line, $\alpha=0.62$) and the linear regression (red-gray line, $\alpha=0.62$) for $n = [10~10^3]$. Again these values were close to the theoretical value $H = 0.625$.}
\label{fig:fig8} 
\end{figure}

With other parameter settings we were also able to obtain persistent behavior, which resembles more closely what has been reported for encephalographic recordings \cite{LinkenkaerHansenetal2001, Palvaetal2013, Botcharovaetal2015, Tonetal2015}. The averaged signal $\bar{X}(t)$ over 1000 realizations with the corresponding DFA results is shown in Fig. \ref{fig:fig8}. Only for the range $n=$[10 10$^3$] our approach indicated power-law behavior with $\alpha=0.62$, which is very close to the expected theoretical value of $H=0.625$; we note that we obtained the same $\alpha$-value using linear regression. 

\subsection{Envelope dynamics of beamformed MEG}
\begin{figure}[h!]
\centering
\subfloat[]{\label{subfig:fig9a}\includegraphics[width=0.32\linewidth]{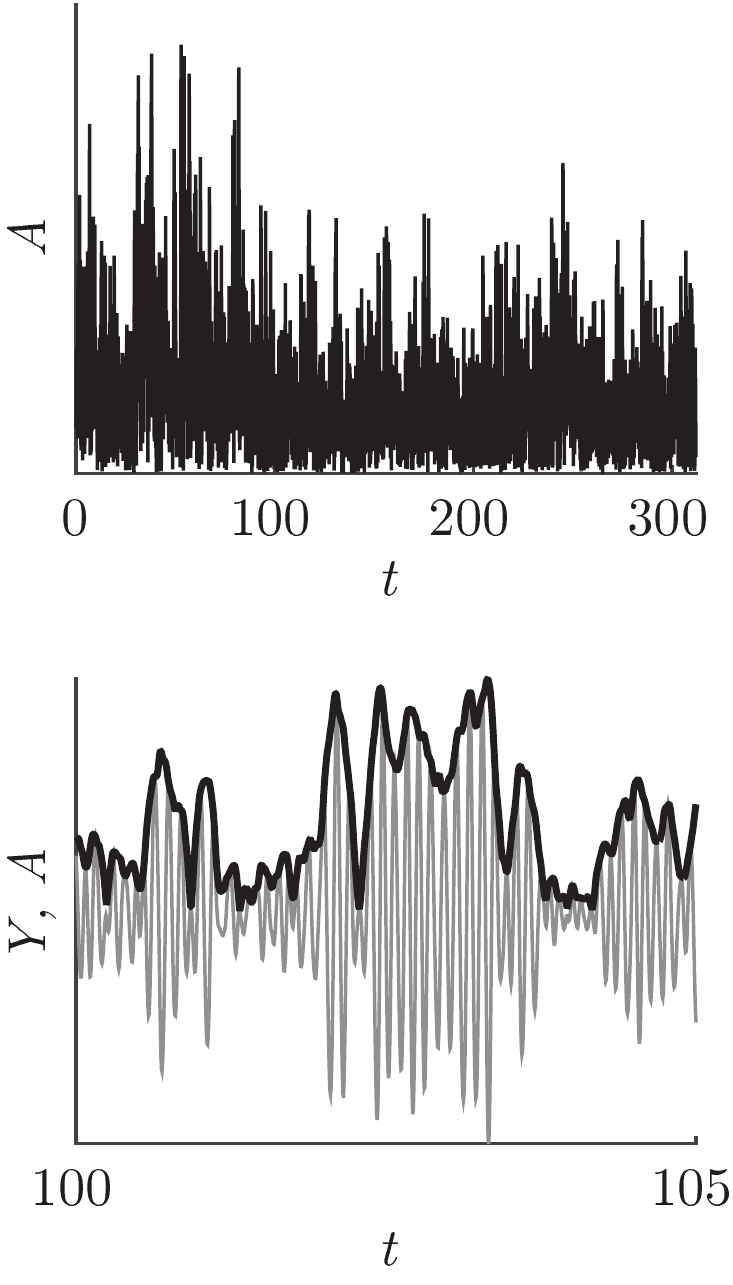}}
\subfloat[]{\label{subfig:fig9b}\includegraphics[width=0.63\linewidth]{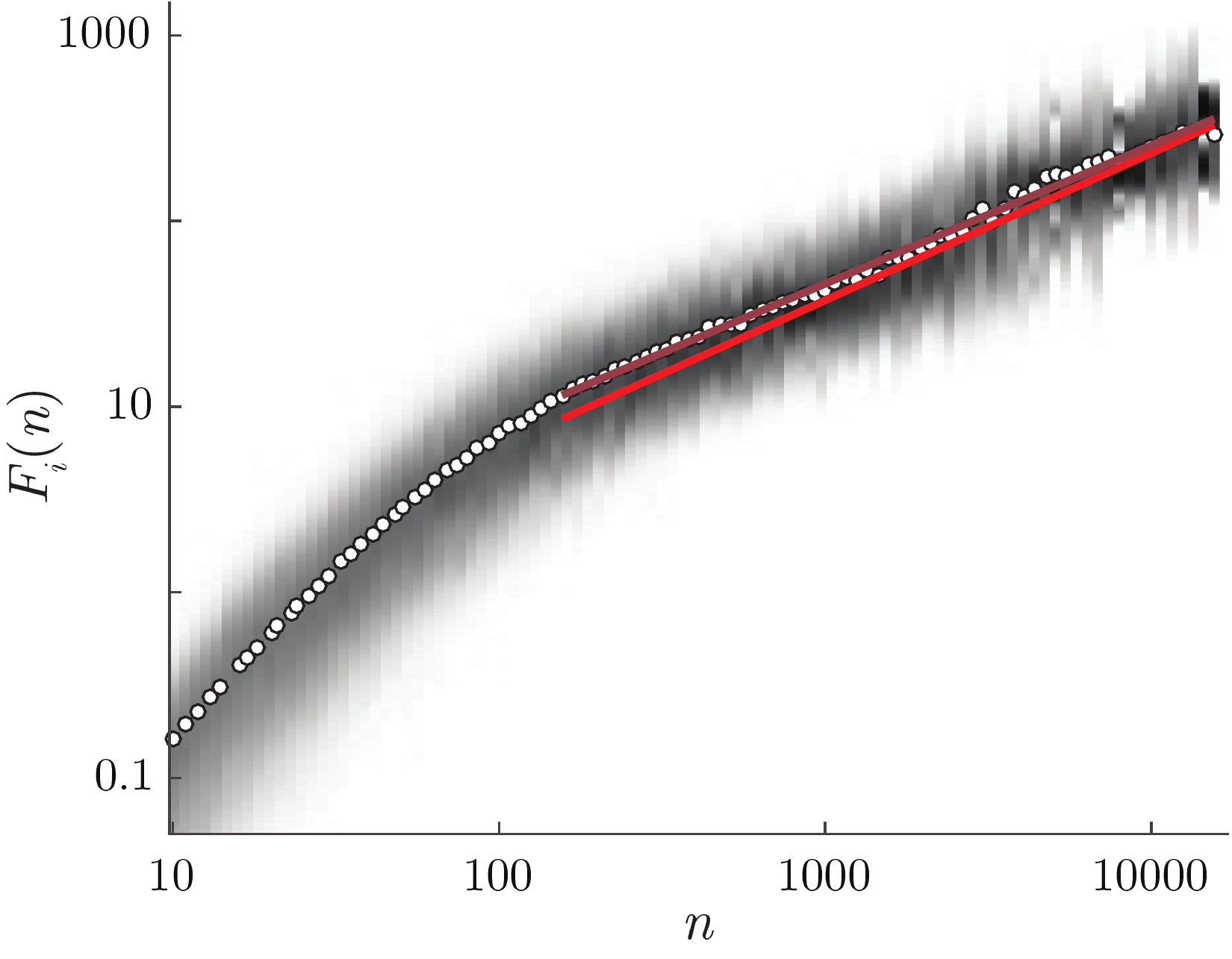}}
\caption{ \small {\bf Fig. \ref{subfig:fig9a}:}  The amplitude envelope $A$ of a single MEG virtual channel signal (right superior occipital gyrus) in the alpha frequency band (8-12 Hz; upper panel). The lower panel shows the filtered signal (gray) together with its envelope $A$ (black); this is a zoom into the full envelope trace shown in the upper panel. {\bf Fig. \ref{subfig:fig9b}:} The distributions $\tilde{p}_n$ (gray shading) together with the maximum likelihood fit (red line, $\alpha=0.79$) and the linear regression (red-gray line, $\alpha=0.75$). Fits were determined over two decades in the range $n = [N/500~N/5]$ with $N = 7.8 \cdot 10^4$ denoting total signal length (312 seconds). For this range both the AIC$_c$ and BIC indicated power-law behavior, in contrast to the results for the entire displayed range of interval sizes $n = [10~N/5]$. The range $n = [N/500~N/5]$ was estimated by first fitting the piecewise-linear model $f_{\theta}^{(10)}$ over the entire range available. This yielded a crossover point of $\log( \theta_4 ) = 2.0$, which (roughly) agreed with the here-chosen $\log (N/500) = 2.2$ as lower end of the range over which power-law scaling was determined. Results for other occipital signals were very similar; see also {\it Appendix B}.}
\label{fig:fig9} 
\end{figure}

\noindent
The results for the alpha amplitude dynamics of a single (virtual) source MEG signal are shown in Fig. \ref{fig:fig9}. In the range $n=[N/500~N/5]$, $N = 7.8 \cdot 10^4 \simeq 312$ seconds, i.e. over two decades, power-law scaling appeared to be present. The range of scaling was predetermined on basis of the crossover point of the piecewise linear model $f^{(10)}_\theta$ fitted over the entire range shown. We note that for that large range ($n=[10~N/5]$) power-law scaling was rejected. In the selected range, the maximum likelihood scaling exponent estimate was $\alpha = 0.79$ (contrasting $\alpha=0.75$ using conventional linear regression). We here report the results for one source only and provide
a more extended re-analysis for all occipital channels and all subjects in {\it Appendix B}; see also \cite{Tonetal2015}.

\section{Discussion}
\noindent
We proposed an approach to assess (the presence of) linearity in the log-log fluctuation plots in DFA. Inspired by \cite{Botcharovaetal2013}, we used a maximum likelihood estimate with the likelihood function $\mathcal{L}$ defined by means of the density functions of the mean-squared fluctuations. This definition allows for using the function $\mathcal{L}$ in assessing power-law behavior and estimating the scaling exponent $\alpha$. 

We estimated $\alpha$ from $\theta_{\text{max}}$, the parameter set that maximizes $\mathcal{L}$. It hence represents the maximum likelihood estimate, rather than the minimal least-squares estimate obtained from linear regression in the conventional DFA approach. Using this estimate one can retrieve the Hurst exponent from fGn processes very accurately. While this may be considered trivial, one has to realize that estimating $\alpha$ by its maximum likelihood estimate is fundamentally different from the standard least squares regression estimate. Our approach uses another notion of an optimal model, unless the fit-residuals are normally distributed \cite{Burnham&Anderson2002}. This approach incorporates the variability in the $F_i$ estimates by means of $p_n$. When variability in $F_i$ increases, this results in a wider density $p_n$. The widening of $p_n$ decreases its contribution to $\mathcal{L}$ by reducing the magnitude of $\tilde{p}_n \left(f_{\theta} \right)$. 

We did not only reliably retrieve scaling exponents in case of proper self-similarity, we could also detect deviations from power-law behavior. We illustrated this by bounding the fluctuation magnitudes using a potential function $U$, leading to dynamic saturation. There, a non-linear function was favored over a linear model to describe the log-log fluctuation magnitude plots, i.e. the hypothesis of power-law behavior had to be rejected. These deviations are not necessarily a result of a deterministic component in the dynamics bounding the fluctuations, but may also stem from mechanisms like periodic trends \cite{Huetal2001}, non-stationarities \cite{Chenetal2002} or non-linear transformations \cite{Chenetal2005}. Finite-size effects caused by taking too small interval sizes may also cause a curvature at these interval sizes \cite{Bryce&Sprague2012}. 

Model selection is a method of comparing models and favoring the models that better describe the available data. Due to its roots in information theory, the comparison between model and data is performed in terms of probability densities. 
In line with the traditional Kullback-Leibner approach, this boils down to evaluating terms like $\int p(x) \ln( q(x|\theta) ) dx$ where $p$ represents the `truth' and $q(x|\theta)$ the probability density originating from an approximating model with parameters $\theta$. 
Interestingly, in the work of Botcharova and co-workers \cite{Botcharovaetal2013} the log-likelihood function was defined as $\ln \left( \mathcal{L}_B \right) = \sum_{n} \log \left( F_s \right)  \ln \left( f_{\theta} \right)$ with $F_s$ denoting a rescaled version of the fluctuation magnitudes $\bar{F}_i$. The fitted model $f_{\theta}$ was normalized such that it integrated to 1. 
While $p$ and $q$ above are probability densities, this is in general not the case for $\mathcal{L}_B$ rendering its interpretation all but trivial. Moreover, due to the rescaling of the fluctuation magnitudes and the normalization of the fitted model, it is impossible to infer parameters, in particular the scaling parameter $\alpha$, from the maximized log-likelihood function in this procedure. As a consequence the optimal model $f_{\theta}$ in the maximum likelihood sense was used to determine the model form \cite{Botcharovaetal2013}, but subsequent estimation of the scaling parameter was obtained through linear regression. That is, the model used in model selection differed from the one used in subsequent parameter estimation. Apparently this is based on a different notion of as to what constitutes an optimal model. Our definition of $\mathcal{L}$ in \eqref{L} does not come with this problem.

We also showed that information criteria tend to favor lower-dimensional models with decreasing $M$ (see Figs. \ref{fig:fig5} and \ref{fig:figA3}) due to the scaling of $\ln \left( \mathcal{L}_{\text{max}} \right)$ with $M$. This is a common phenomenon in model selection \cite{Burnham&Anderson2004, Ahoetal2014} and arises from the fact that model selection targets at minimizing the expected relative directed distance rather than the relative directed distance itself. The reason for this is that $\theta_{\text{max}}$ needs to be estimated as well. Therefore, the estimate $\theta_{\text{max}}$ comes with uncertainty, which increases with the size of the set $\theta_{\text{max}}$, i.e. the number of parameters.
By increasing the sample size $M$, the available evidence increases as well, thus leading to a better estimate of $\theta_{\text{max}}$. For larger $M$ the point at which minimal expected relative distance is reached hence occurs at a larger number of parameters. In the opposite case, the sample size $M$ is too small to provide sufficient evidence for estimating a large set of parameters $\theta$. Therefore the increased uncertainty in $\theta_{\text{max}}$ does not outweigh the gain in goodness-of-fit given by $\mathcal{L}_{\text{max}}$. The effect of $M$ on the magnitude of $\ln \left( \mathcal{L}_{\text{max}} \right)$ resembles the effect of scaling ($F_s$) in $\ln \left( \mathcal{L}_B \right)$ on this same variable. That is, by changing the scaling interval of $F_s$, one may bias the model selection results towards underfitting (smaller interval) or overfitting (larger interval). 

The results in Fig. \ref{fig:fig6} regarding the Kaulakys \& Ruseckas model for neural activity show that the fluctuations $F_i$ resulting from extreme events yield large differences between the averaged values $\bar{F_i}$ and the modes of $p_n$. In consequence the characterization of the autocorrelation structure based on these two measures differed markedly. Averaging signals causes the individual spikes to vanish, which does resemble the superimposed neural contributions to, e.g., encephalographic signals. For the averaged signals the predicted scaling exponents $\alpha$, determined by the parameter values in the system \eqref{Kaulakyseqctu}, can be accurately recovered. We attribute the somewhat limited scaling range of two decades to the reflective boundaries used in \eqref{Kaulakyseqctu}. These results suggest that the signals suitable for DFA should contain symmetry in the distribution of their values around the trends $Y^{\text{trend}}_i(t)$ at least to some degree. Characterization of this symmetry might be obtained by restricting the number or the magnitude of odd cumulants in the generating function \cite{Risken1984}. A detailed discussion of this matter is beyond the scope of the current paper.
We note, however, that studies that considered individual spike trains, e.g, \cite{Fagerholmetal2015, Beggs&Plenz2003,  Shewetal2009, Freyeretal2011} typically employed other outcome variables like the probability distribution. The numerical findings in \cite{Ruseckas&Kaulakys2010} suggest that for individual spike trains, the probability distribution may be more appropriate for characterizing these kinds of processes. 

Our results are largely consistent with the way DFA has been used in neuroscience, especially with regard to $\bar{X}(t)$.
As said, $\bar{X}(t)$ resembles superimposed neural contributions to, e.g., encephalographic signals, which have been the primary target for DFA  \cite{LinkenkaerHansenetal2001,Palvaetal2013,Botcharovaetal2015}. We recently applied our approach to MEG data \cite{Tonetal2015} and sketched a part of this analysis in Fig. \ref{fig:fig9}. While this illustration already shows the feasibility of our approach in the context of neurophysiological data, we would like to add that the study of MEG signals in \cite{Tonetal2015} used a separation of time scales of the underlying neural dynamics as a starting point \cite{Haken1977}. Accordingly, we sought to identify order parameters capturing the dynamics of whole brain activity. Without affecting the individual scaling characteristics, we z-scored the order parameter time series for each subject, determined $p_n$ on basis of the pooled results, and demonstrated their self-similarity. 

The question which information criterion should be used in model selection, in this case AIC$_c$ or BIC, cannot be unambiguously given \cite{Ahoetal2014}. This is because it is ultimately a philosophical question, as it depends on whether we expect the `truth' to be contained in our candidate model set. In general one may distinguish two scenarios \cite{Ahoetal2014}: The first is a very complex model producing the data, such that one does not expect the sample size to exceed the number of parameters in this model, in our case amounting to $M \ll K$. That is, one does not expect the correct model, i.e. the model equalling the `truth', to be contained in the set of candidate models. In this case the objective is to find the model with optimal accuracy in describing the data. In the second scenario the data set is generated by a relatively simple process: $M \gg K$. Because the model is comparably simple, one may assume that it equals one of the models in the candidate model set and one thus aims for finding the `true' model. 
Thus, the assumption regarding data complexity changes the model selection objective, where AIC$_c$ is more appropriate in the first case and BIC in the second. But which of the two sketched scenarios applies here?
Botcharova et al. \cite{Botcharovaetal2013} advised to use AIC$_c$, because it may lead to fewer false positives and more reliably captures the form of the fluctuation plot when periodicities are present in the underlying signal. Contrasting this somewhat heuristic argument, the conceptual framework of synergetics \cite{Haken1977} suggests that the dynamics of complex systems like the human brain may be captured by a 'simple' model: In the vicinity of non-equilibrium phase transitions a complex system admits a low-dimensional, hence 'simple' dynamics. Combining this with the emerging idea that the brain resembles a system in a permanently critical state \cite{Shew&Plenz2013}, this could suggest that the BIC may be more suited for characterizing dynamics in brain activity.
Obviously, the question about which information criterion to use is still a matter of debate. We do not indicate any preference. 

The approach discussed in this paper is not restricted to DFA but applies to a variety of situations aimed at finding some relationship between two variables with multiple estimates for each value of the independent variable. This could either be real-world data where many measurements are feasible, but also simulation studies applying a Monte-Carlo scheme. Its roots in likelihood theory renders the current procedure naturally suited to stochastic systems. 

\section*{Acknowledgement}
\noindent
This work was funded by the ERC Advanced Grant DYSTRUCTURE (n. 295129). We thank Mark Woolrich and Morten Kringelbach for collecting and pre-processing the MEG data.

\newpage

\newpage
\begin{appendices}
\renewcommand{\theequation}{\Alph{section}.\arabic{equation}}
\numberwithin{equation}{section}
\renewcommand{\thefigure}{\Alph{section}.\arabic{figure}}
\numberwithin{figure}{section}
\numberwithin{table}{section}

\section{Appendix A}
\subsection{Toy model for neural mass dynamics}
\noindent
We consider a class of stochastic systems proposed by \cite{Ruseckas&Kaulakys2010}, which display $1/f$ power spectra. That structure is induced by the multiplicative noise term in the SDE governing its dynamics:
\begin{align}
d X(t) = \sigma^2 \left[ \eta - \frac{\lambda}{2} \right] X(t)^{2 \eta -1} + \sigma X(t)^{\eta} dB(t) \label{Kaulakyseqctu}
\end{align}
The derivation of \eqref{Kaulakyseqctu} is based on a point process model with stochastic inter-pulse intervals. The type of behavior of these models bears resemblance with neuronal spike trains, which motivated considering them here. Numerical integration was implemented using the following discretization of the dynamics \eqref{Kaulakyseqctu}:
\begin{align}
\begin{split}
X_{k+1} &=  \left( 1 + \frac{\eta - \lambda}{2} \kappa^2 \right) X_k + \kappa X_k \xi_k \\
t_{k+1} &= t_k + \left( \frac{\kappa}{\sigma} \right)^2 X_k^{2(1-\eta)} 
\end{split} \label{Kaulakyseqdisc}
\end{align}
We used variable time steps for the integration to cover the temporal behavior of the spikes; $\xi_k$ denotes a conventional Gaussian white noise process. To limit the diffusion of $X(t)$ we employed reflective boundaries at $X_{\text{min}} = 1$ and $X_{\text{max}} = 10^6$. All simulations were performed until $t=1000$ was reached and subsequently resampled to obtain an equally spaced time axis of length $N=10^6$. This resampling is mandatory for a proper interpretation of the DFA results.

We ran simulations with two different sets of parameters: $\left\{ \sigma=1, \eta = 2, \lambda = 4 \right\}$ and $\left\{ \sigma=2, \eta = 0.4, \lambda = 1.5 \right\}$ with $\kappa = 0.1$ for both cases. Following \cite{Ruseckas&Kaulakys2010}, the exponent $\beta$ in the power spectral density $S(\omega) = \omega^{-\beta}$, relates to the parameters $\lambda$ and $\eta$ as $\beta = 1 + (\lambda-3)/(2(\eta-1))$. When using $1+2H = \beta$, which holds for a fractional Brownian motion process \cite{Daffertshofer2011}, we obtain
\begin{align}
H = \frac{\lambda-3}{4(\eta-1)} \label{Hurstrelation}
\end{align}
as the relation between parameter values and the Hurst exponent. From this relation it follows that $\left\{ \sigma=1, \eta = 2, \lambda = 4 \right\}$ corresponds to an anti-persistent process with $H=0.25$, whereas $\left\{ \sigma=2, \eta = 0.4, \lambda = 1.5 \right\}$ leads to a persistent process with Hurst exponent $H=0.625$. In \cite{Ruseckas&Kaulakys2010} the exponent $\beta$ in the power spectrum was not allowed to exceed 2 suggesting an upper bound $H=\frac{\beta-1}{2} = 0.5$. This would imply that only anti-persistent processes 
could be generated by \eqref{Kaulakyseqctu}. Here, however, we show in Fig. \ref{fig:fig8} that a persistent process can also be generated albeit for a limited scaling range $n = [10~10^3]$. 

\subsection{AIC$_c$ results}
\noindent
\begin{figure}[h!]
\centering 
\includegraphics[width=0.75\textwidth]{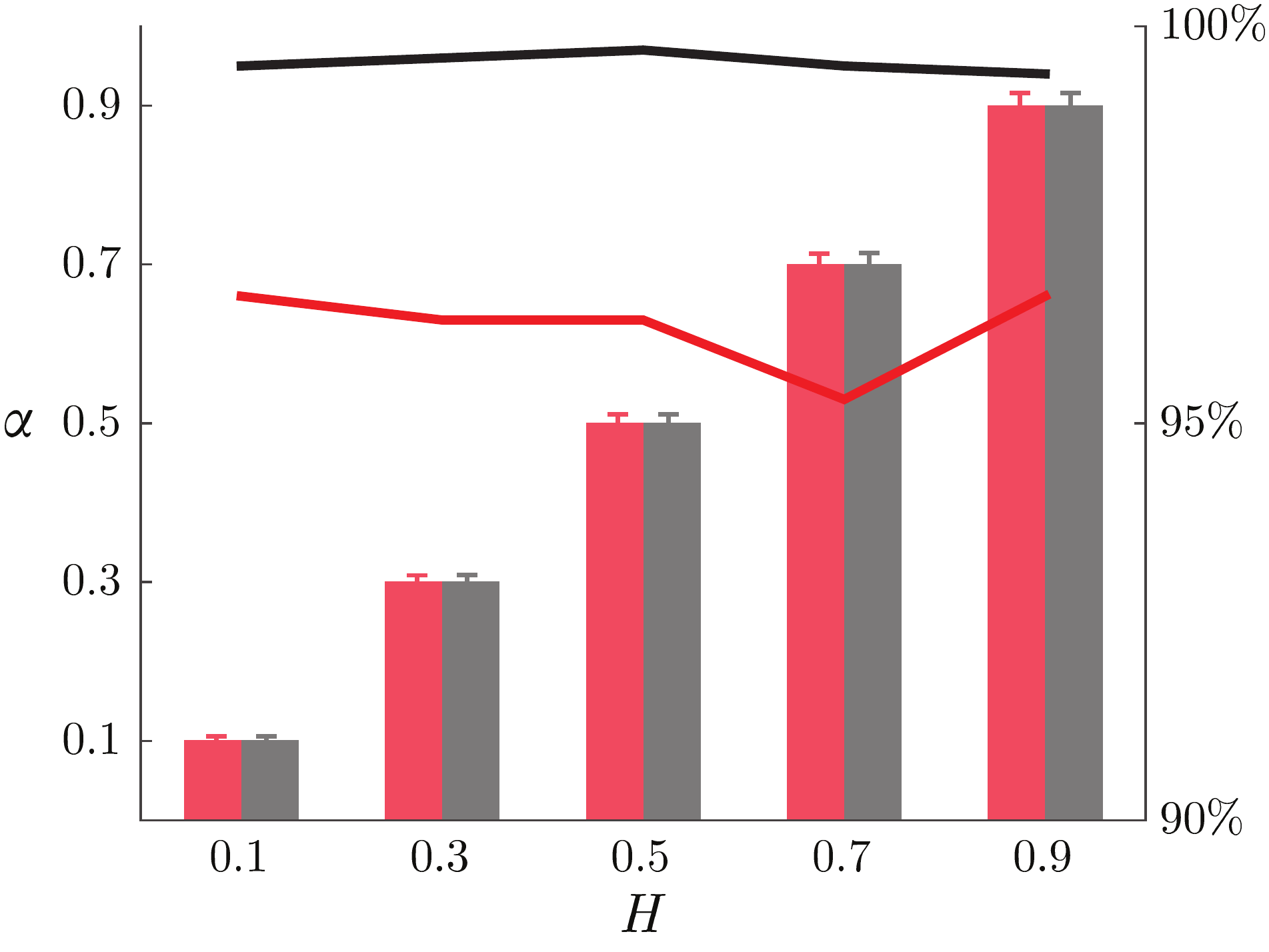}
\caption{\small {\bf Fig. \ref{fig:figA1}:} Performance for the $\Delta B_H(t)$ process for the values $H$ depicted on the horizontal axis. Proportion of the cases in which the linear model was preferred is on the right vertical axis and denoted by the black (BIC) and red (AIC$_c$) lines. On the left vertical axis $\alpha$ represents the averaged scaling exponent estimate over all realizations that resulted in a preferred linear model in case AIC$_c$ (red bars) or BIC (black bars) was used as a criterion. Absolute standard deviations are given by the error bars. } 
\label{fig:figA1}
\end{figure}
Fig. \ref{fig:figA1} is the equivalent of Fig. \ref{fig:fig3} but here the results are displayed for both the AIC$_c$ and BIC and shows the qualitative similarity between these results. The proportion of realizations for which the linear model was favored amounted to [96.6\%, 96.3\%, 96.3\%, 95.3\%, 96.6\%] (AIC$_c$) and [99.5\%, 99.6\%, 99.7\%, 99.5\%, 99.4\%] (BIC) for $H =$ [0.1, 0.3, 0.5, 0.7, 0.9] respectively. The relative errors $ \frac{H - \alpha}{H}$ were [1.3\%, 0.3\%, 0.1\%, 0.1\%, 0.1\%] for the AIC$_c$ and [1.1\%, 0.3\%, 0.1\%, 0.1\%, 0.1\%] for the BIC with relative standard deviations [4.1\%, 2.6\%, 2.1\%, 1.8\%, 1.7\%] (AIC$_c$) and [4.2\%, 2.7\%, 2.2\%, 1.9\%, 1.7\%] (BIC). We thus conclude that in the case of self-similar signals in the form of a fGn, both information criteria indicate a power-law in the large majority of realizations with an very accurate scaling exponent estimate. The (minor) differences between the estimates using AIC$_c$ and BIC are caused by the fact that the results only contain those realizations that were classified as being a power-law. The $\alpha$ estimate for one realization was not affected by the choice of criterion. 
\begin{figure}[h!]
\centering
\includegraphics[width = 0.75\textwidth]{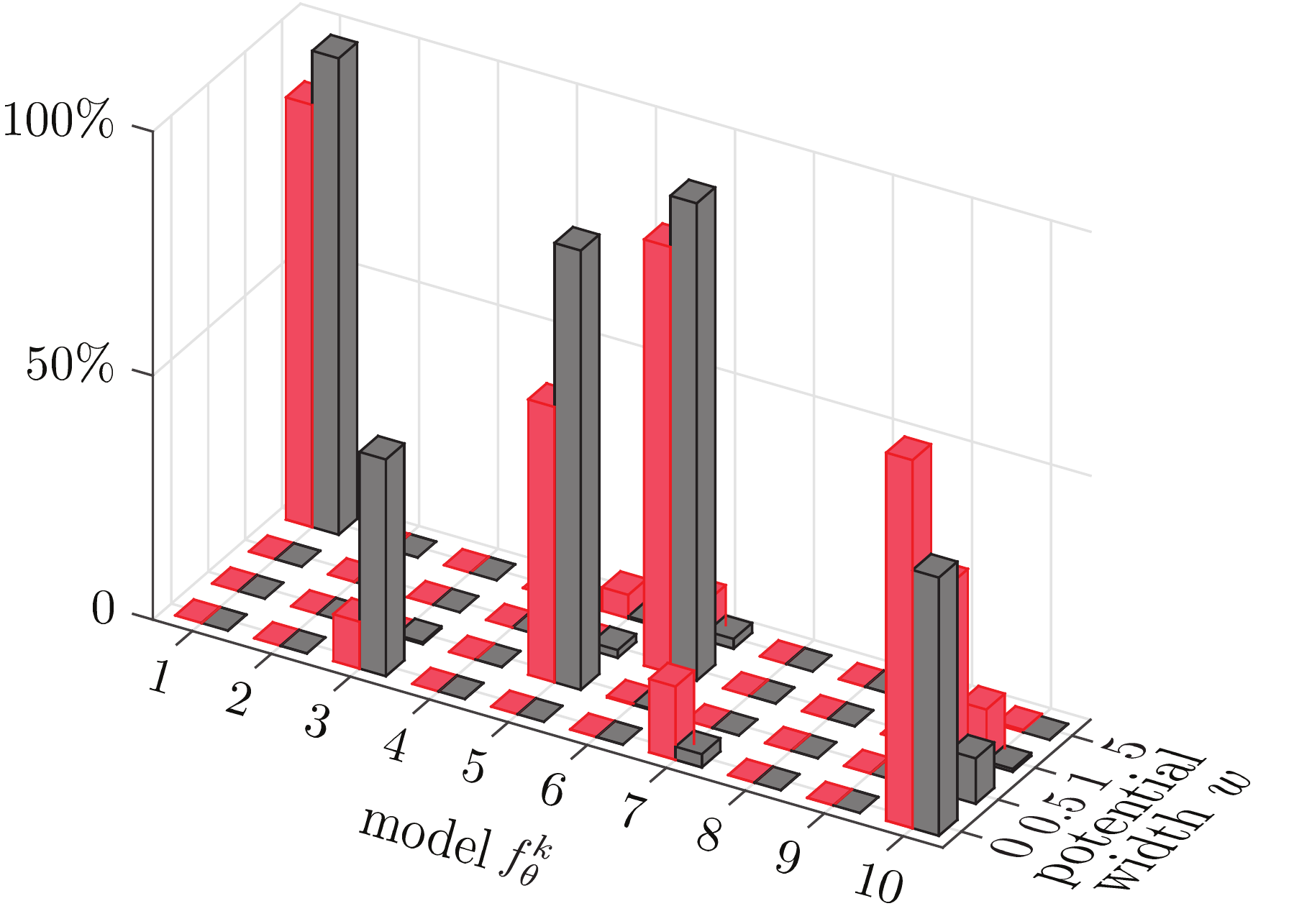} 
\caption{\small {\bf Fig. \ref{fig:figA2}:} Proportions of the total number of 1000 realizations for which the indicated model was preferred in case of AIC$_c$ (red) and BIC (black) for 4 different $w$ (potential width) values. Numbers on the non-labeled axis correspond to different forms of the model $f_{\theta}$ defined in \eqref{models1}-\eqref{models4}. The Hurst exponent of the noise process equalled $1/2$ }
\label{fig:figA2} 
\end{figure}
\noindent
Fig. \ref{fig:figA1}, Fig. \ref{fig:figA2} is the counterpart of Fig. \ref{fig:fig4}. For the $w=0$ case, three different non-linear models were assigned to the data:
 $f_{\theta}^{3}$ (1.0\%, 44.4\%), $f_{\theta}^{7}$ (15.0\%, 0.3\%), and $f_{\theta}^{10}$ (75.4\%, 52.9\%) with percentages indicating AIC$_c$ and BIC, respectively. Only when increasing the potential width to $w = 5$, the employed range of interval sizes was insufficiently large to reveal the bounding effect of the potential. This led to assigning the linear model $f_{\theta}^1$ in 86.5\% (AIC$_c$) and 97.6\% (BIC) of all realizations.
\begin{figure}[h!]
\centering
\subfloat[]{\label{subfig:figA3a}\includegraphics[width=0.32\textwidth]{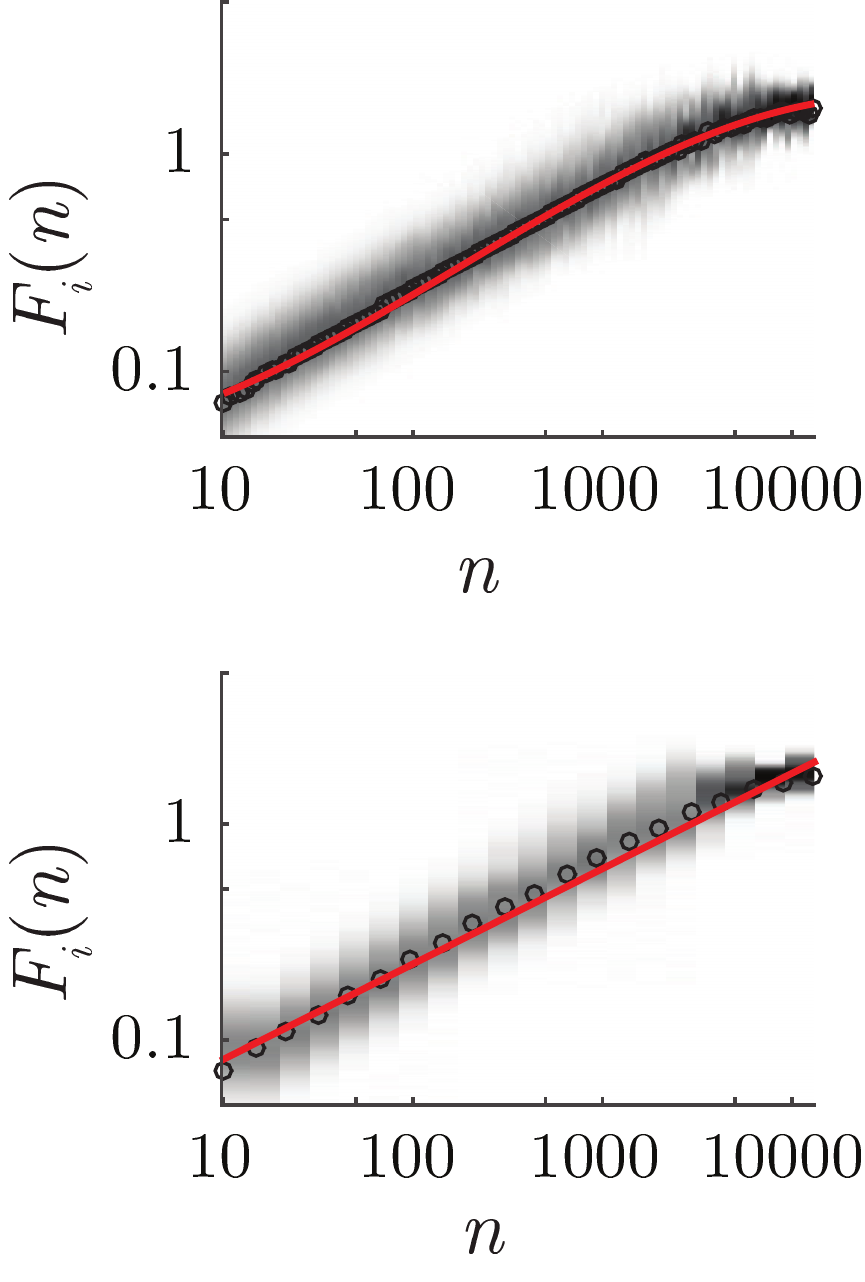}}
\subfloat[]{\label{subfig:figA3b}\includegraphics[width = 0.63\textwidth]{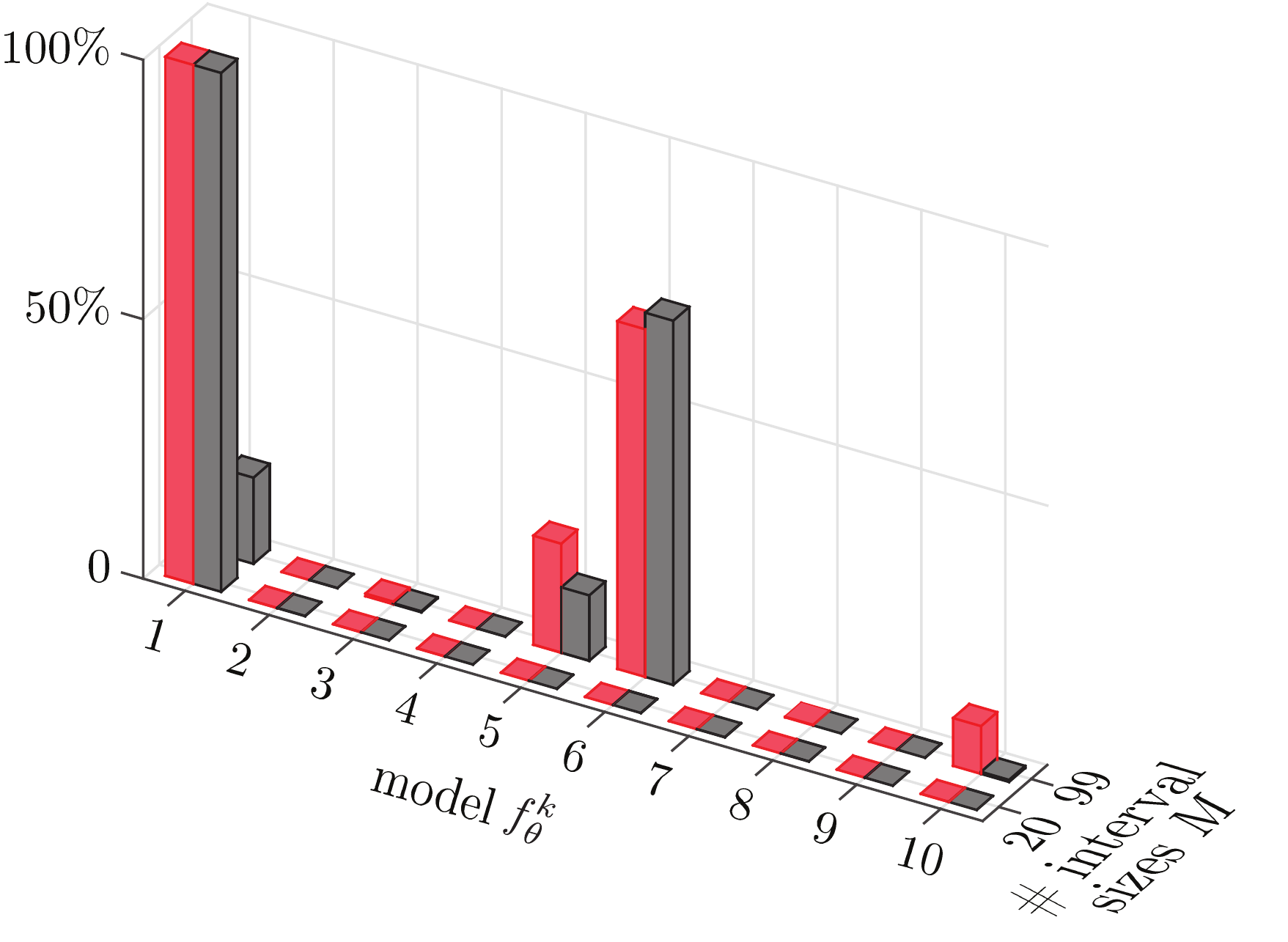}}
\caption{ \small {\bf Fig. \ref{subfig:figA3a}:} Typical DFA results of a process given by \eqref{fGnPotSDE} with $w = 2.5$ and $H=\frac{1}{2}$. Colors represent $p_n$ values, black circles indicate $\left \langle \tilde{F}_i(n) \right \rangle$ values. The optimal fits (red lines) were determined by AIC$_c$ (but BIC agreed for this realization). For $M = 99$ model $f_{\theta}^6(x)$ was selected (upper panel), for $M= 20$ the linear fit $f_{\theta}^1(x)$ (lower panel). {\bf Fig. \ref{subfig:figA3b}:} Similar to Fig. \ref{fig:fig5}. Proportions of the total number of 1000 realizations for which the indicated model was preferred in case of AIC$_c$ (red) and BIC (black) as function of the number of interval sizes $M$. The same realizations were used for both $M$ values. Numbers on the $f_{\theta}^k$ axis correspond to different forms of the model $f_{\theta}$ defined in \eqref{models1}-\eqref{models4}. }
\label{fig:figA3} 
\end{figure}

Fig. \ref{fig:figA3} summarizes the effect of number of interval sizes $M$ on model selection. The results are based on sample paths generated by \eqref{fGnPotSDE} with $w = 2.5$. The linear model was preferred in all cases for $M = 20$. For the default $M = 99$, the increased evidence for non-linearity in the log-log fluctuation plot led to a rejection of the linear model: only 0.2\% (AIC$_c$) and 16.6\% (BIC) of the realizations were classified as a power law. Again we found that BIC tended to favor a lower-dimensional model, but qualitatively results agreed across criteria.

\section{Appendix B}
%
\noindent
We present a brief re-analysis of MEG recordings discussed in conjunction with Fig. \ref{fig:fig9} in the main text. For more details on data acquisition and pre-processing we refer to \cite{Cabraletal2014,Tonetal2015}.

Signals were filtered with an second order IIR bandpass filter into five distinct frequency bands: delta (2-4 Hz), theta (4-8 Hz), alpha (8-12 Hz), beta (20-30 Hz) and gamma (40-80 Hz). We selected a range of interval sizes $n = [N/500~N/5]$, with $N = 7.3 \cdot 10^4 \simeq 292$ seconds corresponding to the number of samples in the shortest time series available. We determined the $\alpha$ scaling exponents using conventional DFA and using the here proposed method for each subject and each source yielding $\bar{\alpha}$ and $\alpha_{\text{AIC}}/\alpha_{\text{BIC}}$, respectively | see Tab. \ref{tab:tabB1}. Scaling exponents were averaged over all 60 signals ($\bar{\alpha}$) or over those signals that were classified as a power law ($\alpha_{\text{AIC}}/\alpha_{\text{BIC}}$). We further listed the corresponding proportions of the total number of signals, i.e. $P_{\text{AIC}}$ and $P_{\text{BIC}}$, respectively. For these same signals, i.e. the ones classified as a power law, we also averaged the conventional estimates whose values are denoted by $\bar{\alpha}_{\text{AIC}}$ and $\bar{\alpha}_{\text{BIC}}$, respectively.

\begin{figure}[h!]
\centering
\subfloat[]{\label{subfig:figS1a}\includegraphics[width=0.5\linewidth]{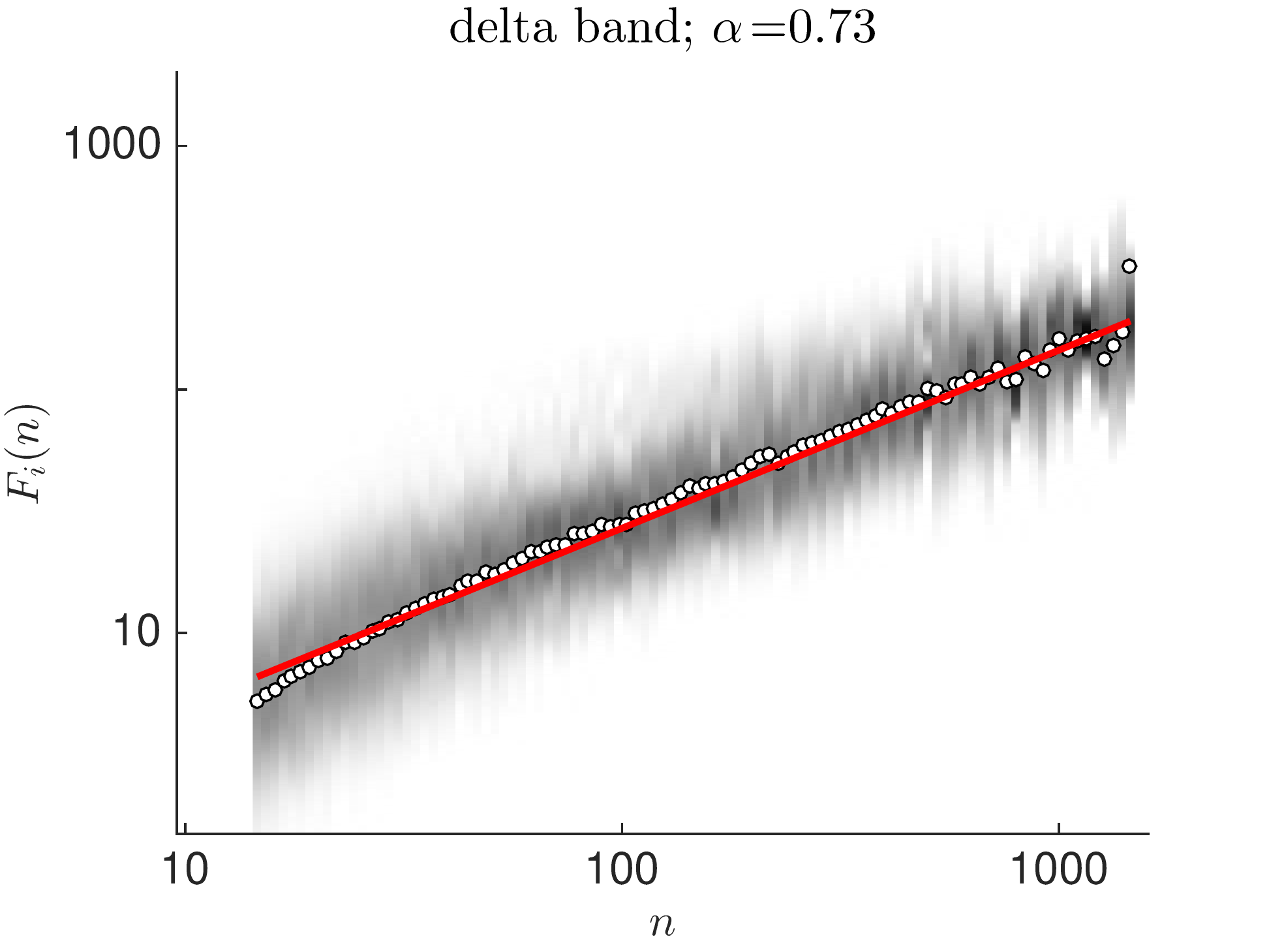}}
\subfloat[]{\label{subfig:figS1b}\includegraphics[width=0.5\linewidth]{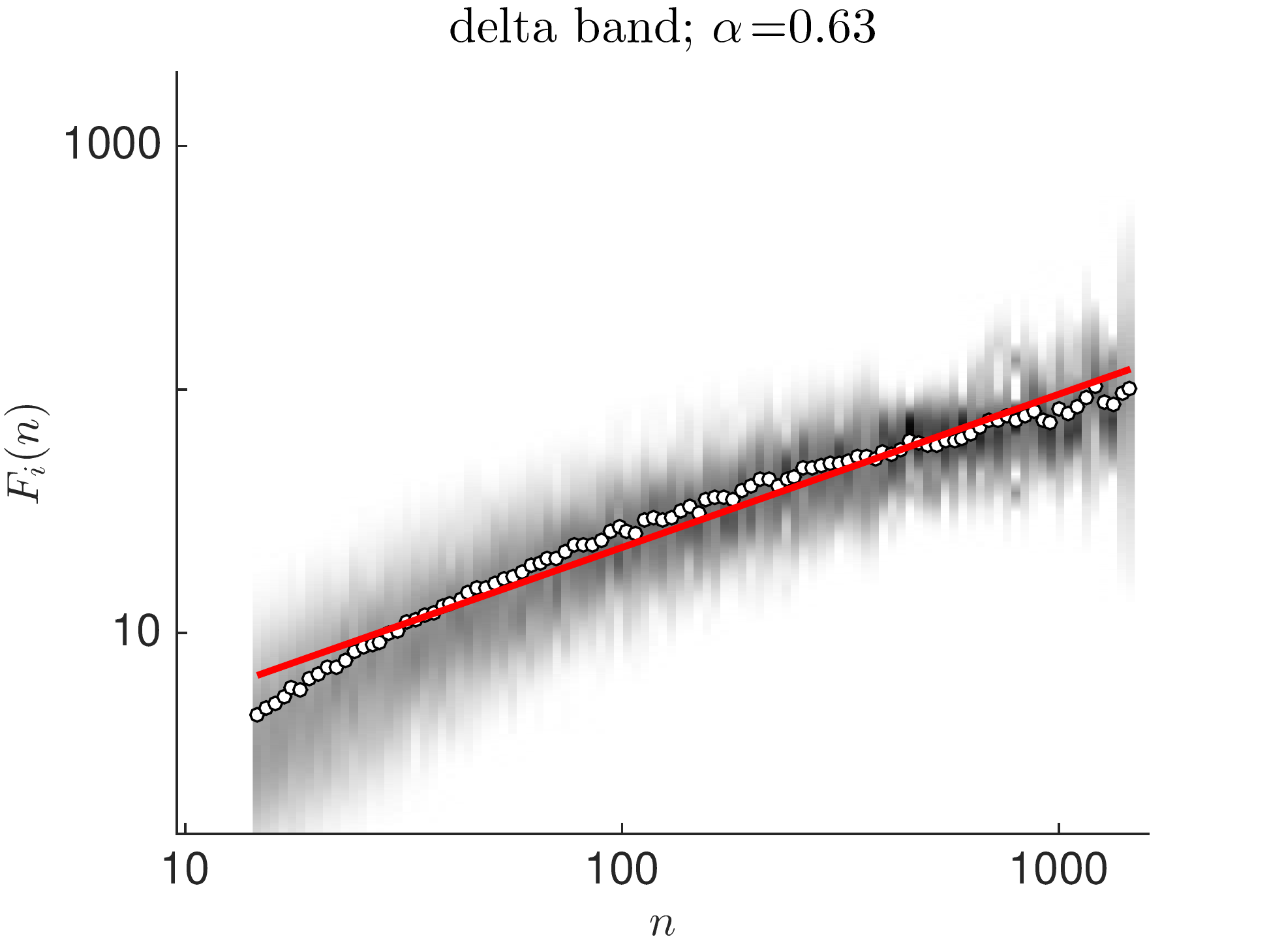}}
\caption{ \small {\bf Fig. \ref{fig:figS1}:} Examples of DFA results in the delta band for a signal with a power law autocorrelation function (Fig. \ref{subfig:figS1a}) and for one without (Fig. \ref{subfig:figS1b}). The circles represent the expected values $\mathbb{E}\left[\tilde{F}_i\right]$ corresponding to the distributions $\tilde{p}_n$ (see \eqref{EF}) in Fig. \ref{subfig:figS1a} or the averaged values $\bar{F}_i$ in Fig. \ref{subfig:figS1b}. The red lines correspond to the maximum likelihood fit in case the fluctuation plot was classified as a power law (Fig. \ref{subfig:figS1a}) or the conventional least squares fit when a power law was rejected (Fig. \ref{subfig:figS1b}).} \label{fig:figS1}
\end{figure}

\begin{figure}
\centering
\subfloat[]{\label{subfig:figS2a}\includegraphics[width=0.48\linewidth]{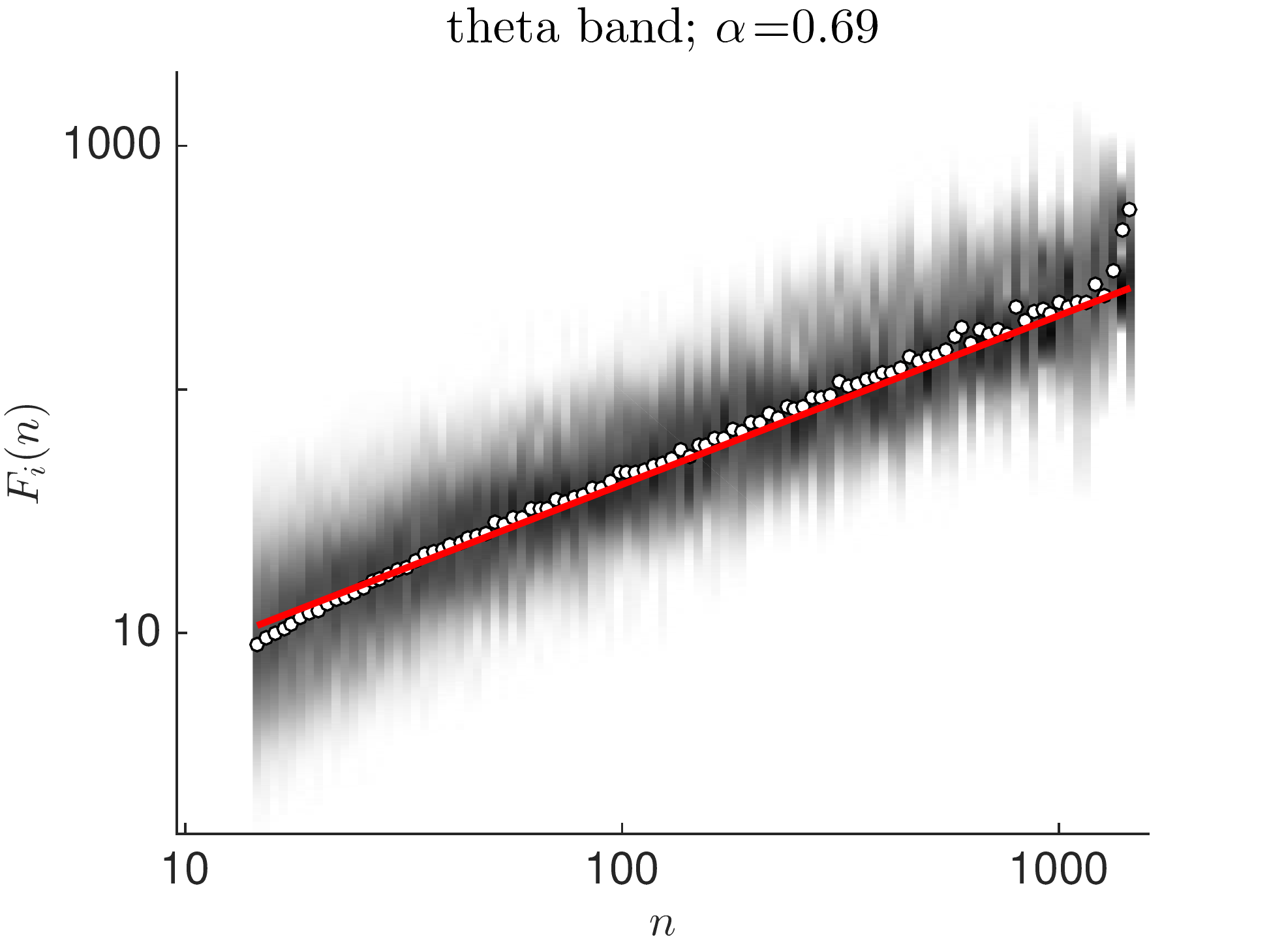}}
\subfloat[]{\label{subfig:figS2b}\includegraphics[width=0.48\linewidth]{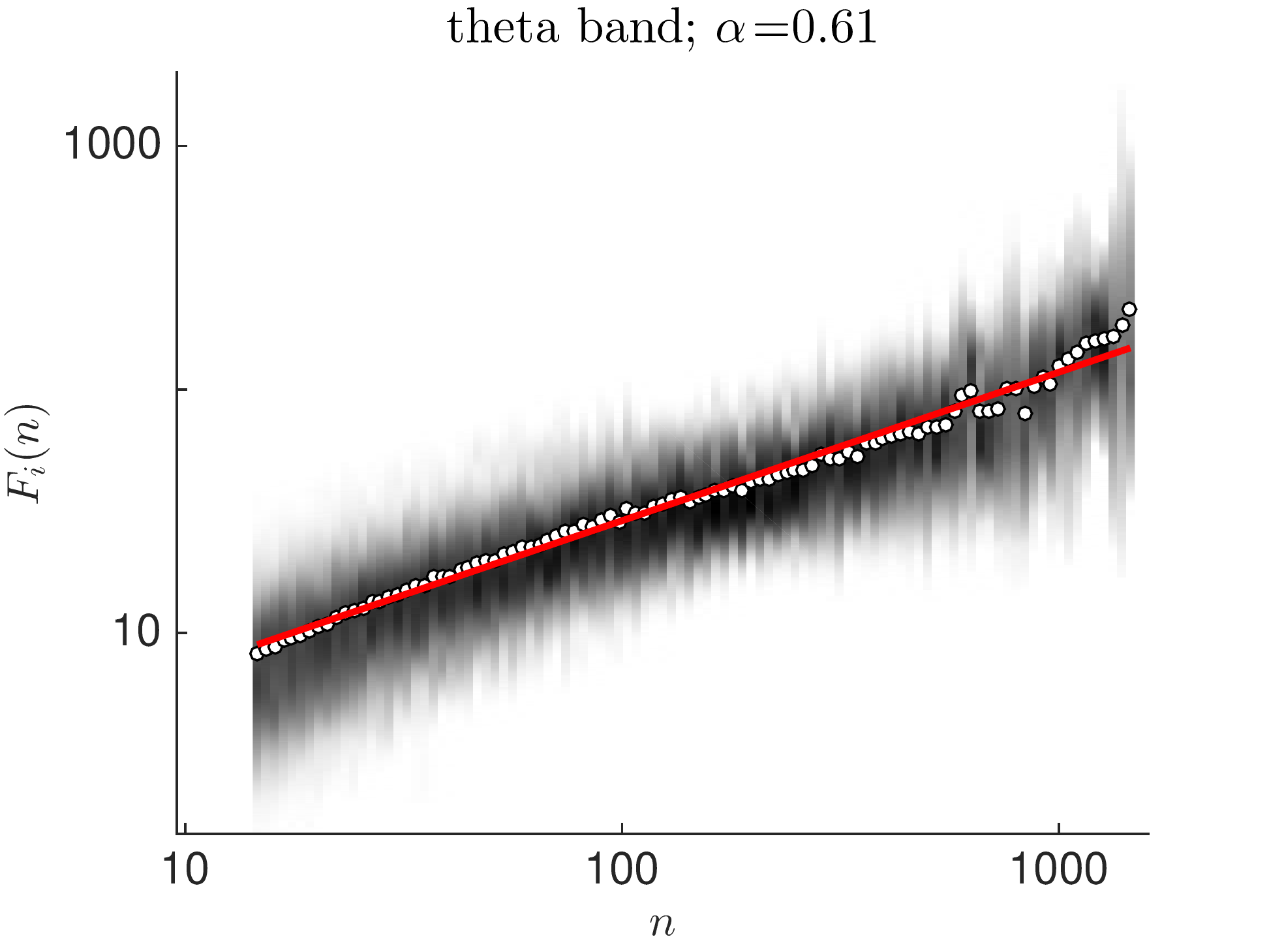}}
\caption{ \small {\bf Fig. \ref{fig:figS2}:} Idem as Fig. \ref{fig:figS1} but for the theta band.} \label{fig:figS2}
\end{figure}

\begin{figure}
\centering
\subfloat[]{\label{subfig:figS3a}\includegraphics[width=0.48\linewidth]{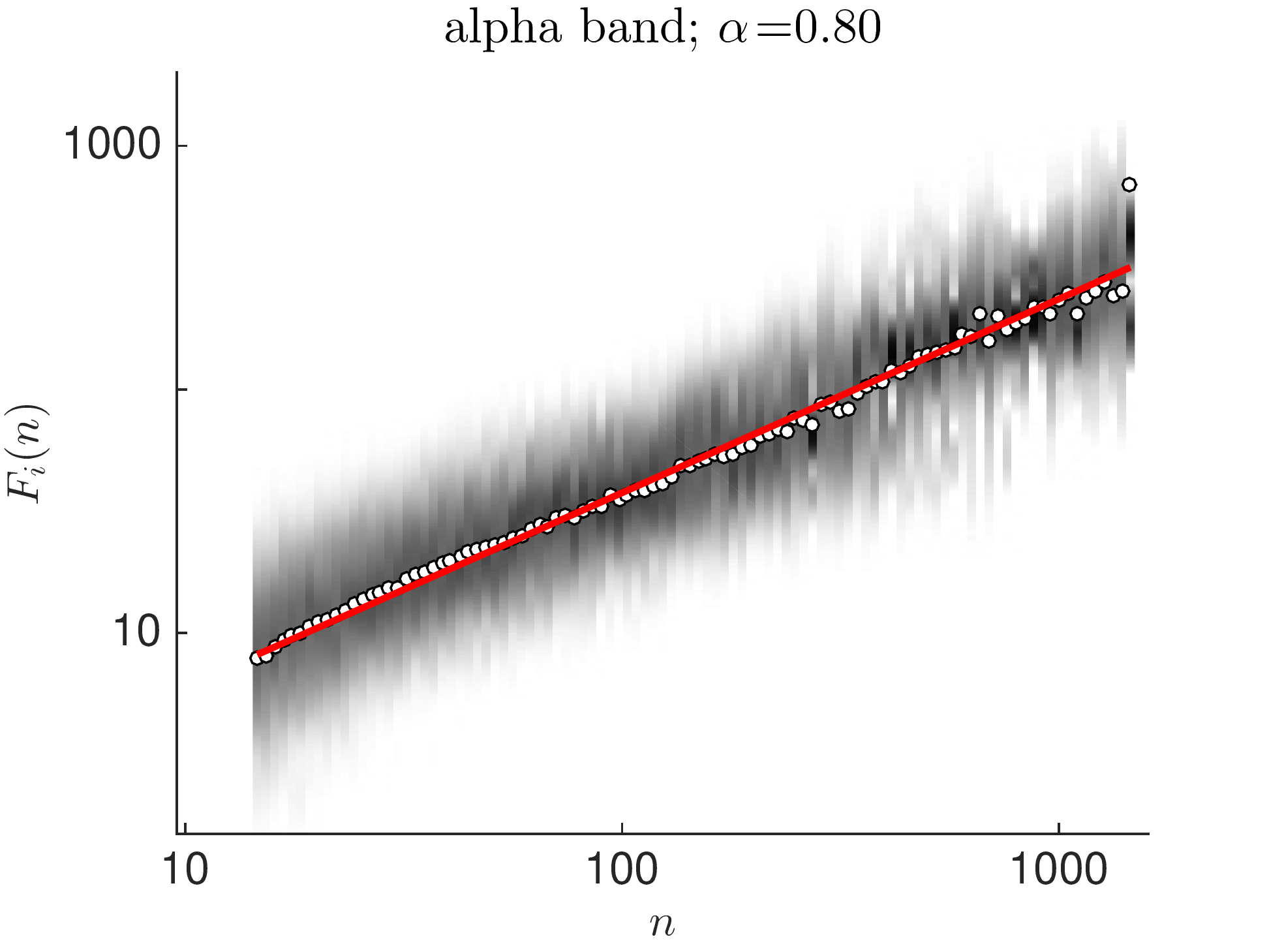}}
\subfloat[]{\label{subfig:figS3b}\includegraphics[width=0.48\linewidth]{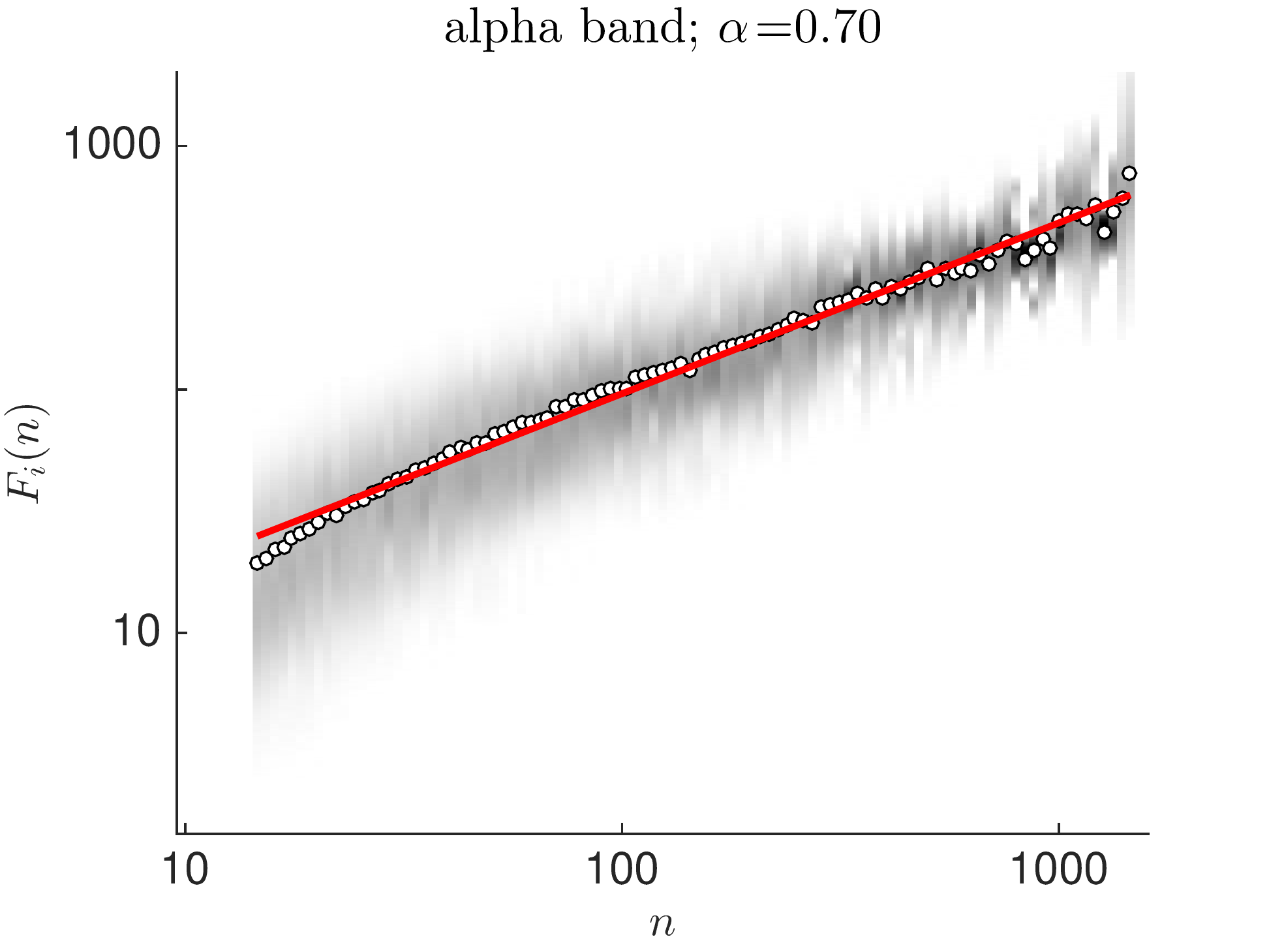}}
\caption{ \small {\bf Fig. \ref{fig:figS3}:} Idem as Fig. \ref{fig:figS1} but for the alpha band.} \label{fig:figS3}
\end{figure}

\begin{figure}
\centering
\subfloat[]{\label{subfig:figS4a}\includegraphics[width=0.48\linewidth]{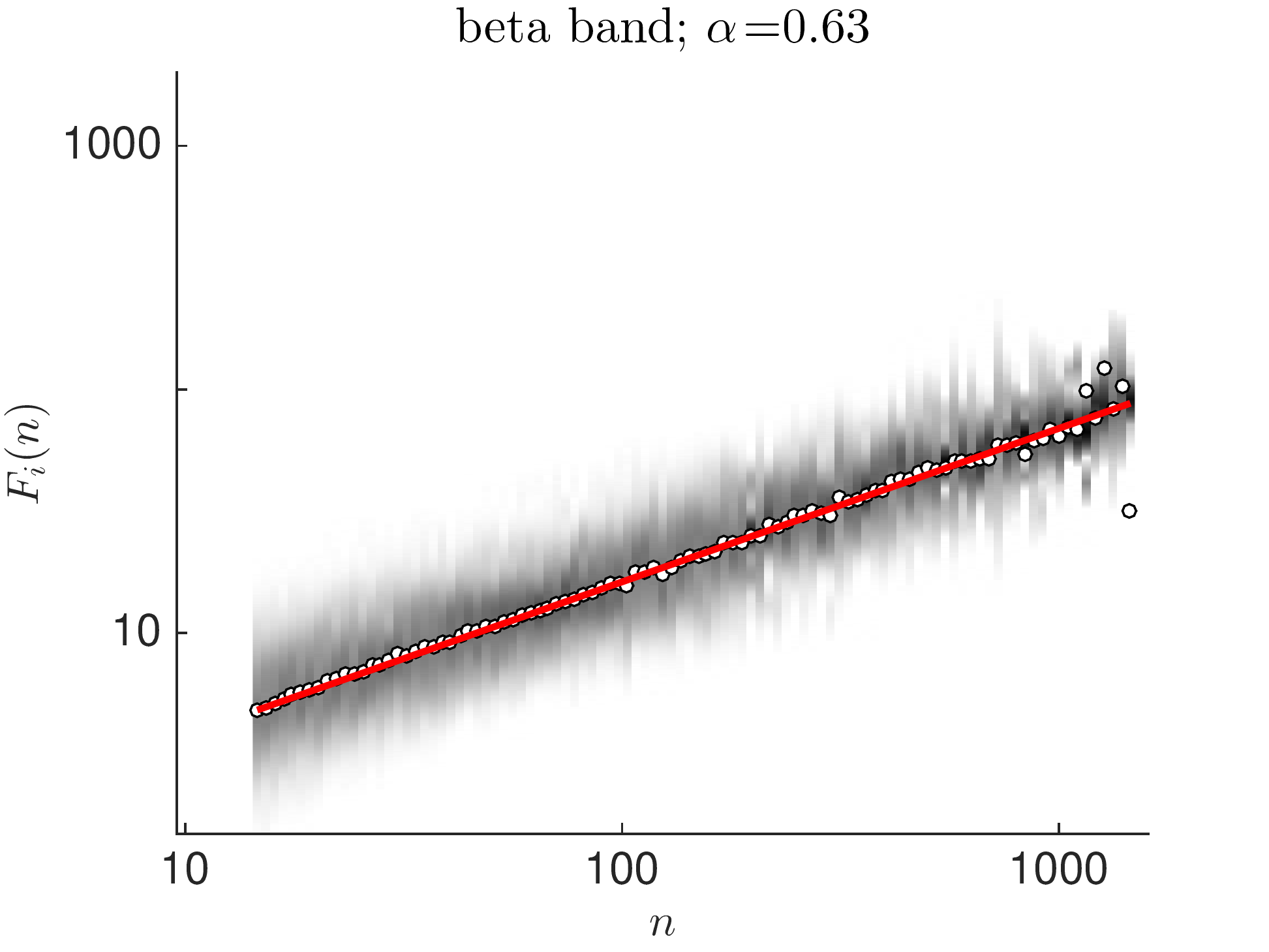}}
\subfloat[]{\label{subfig:figS4b}\includegraphics[width=0.48\linewidth]{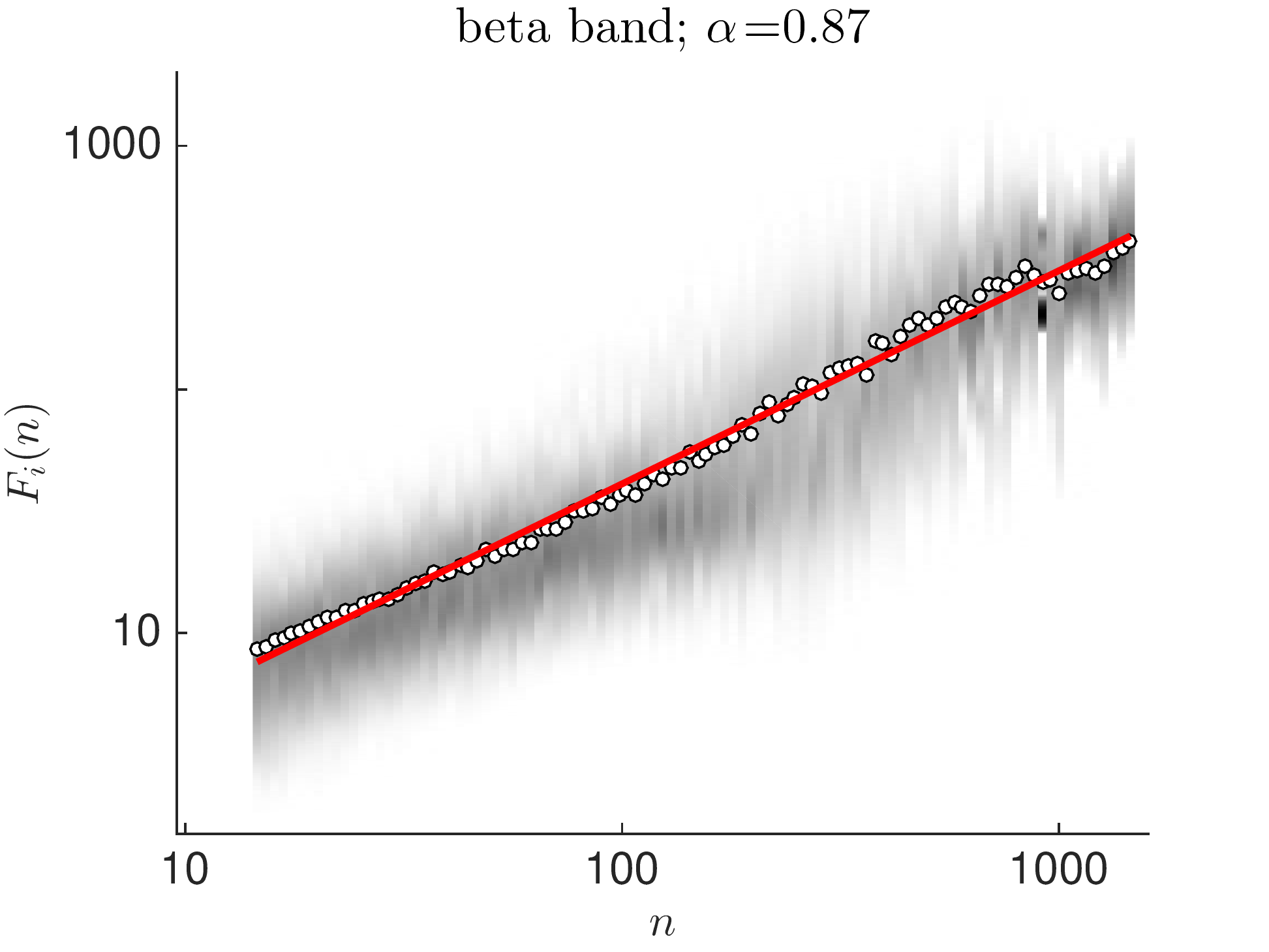}}
\caption{ \small {\bf Fig. \ref{fig:figS4}:} Idem as Fig. \ref{fig:figS1} but for the beta band.} \label{fig:figS4}
\end{figure}

\begin{figure}
\centering
\subfloat[]{\label{subfig:figS5a}\includegraphics[width=0.48\linewidth]{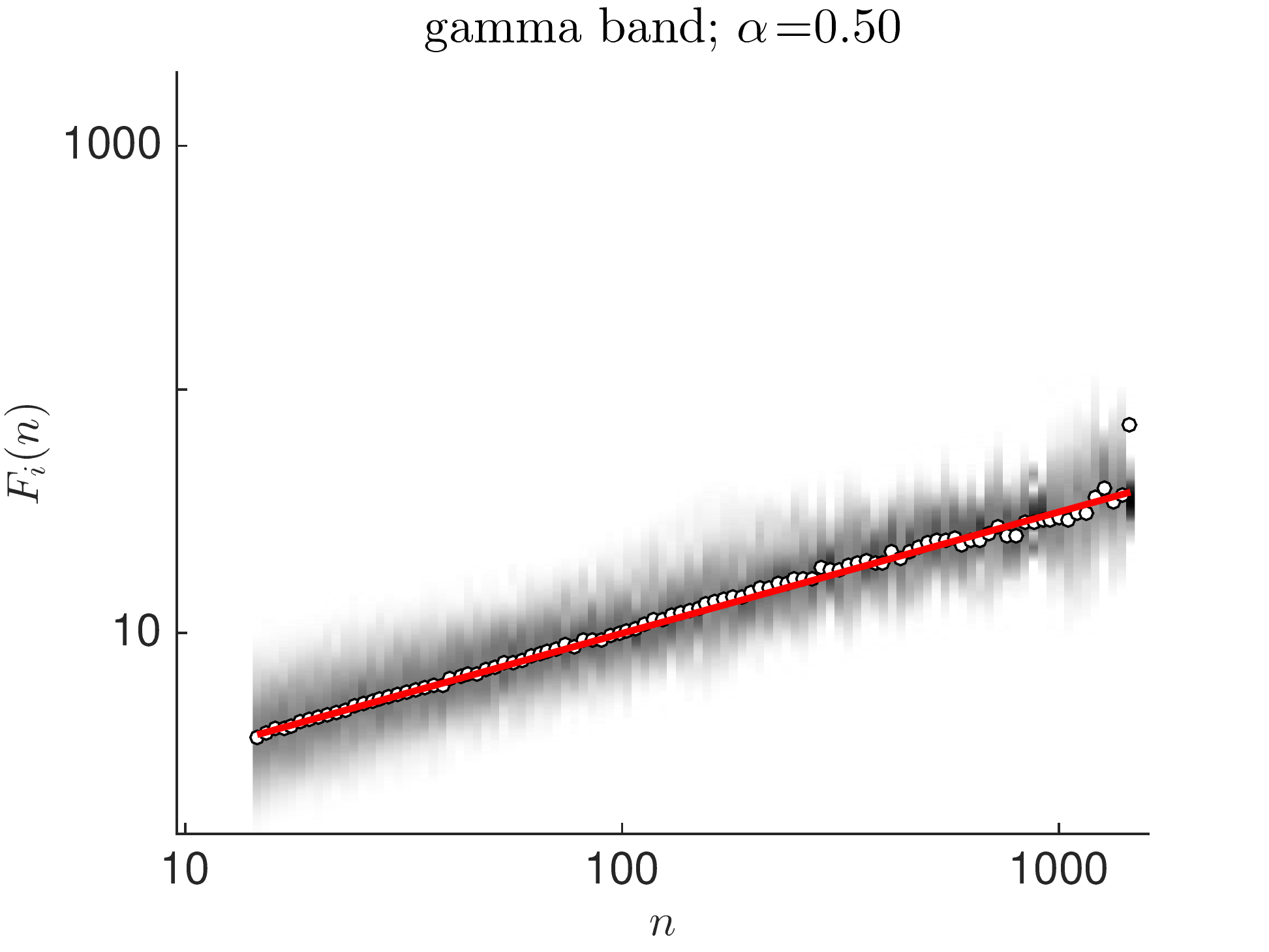}}
\subfloat[]{\label{subfig:figS5b}\includegraphics[width=0.48\linewidth]{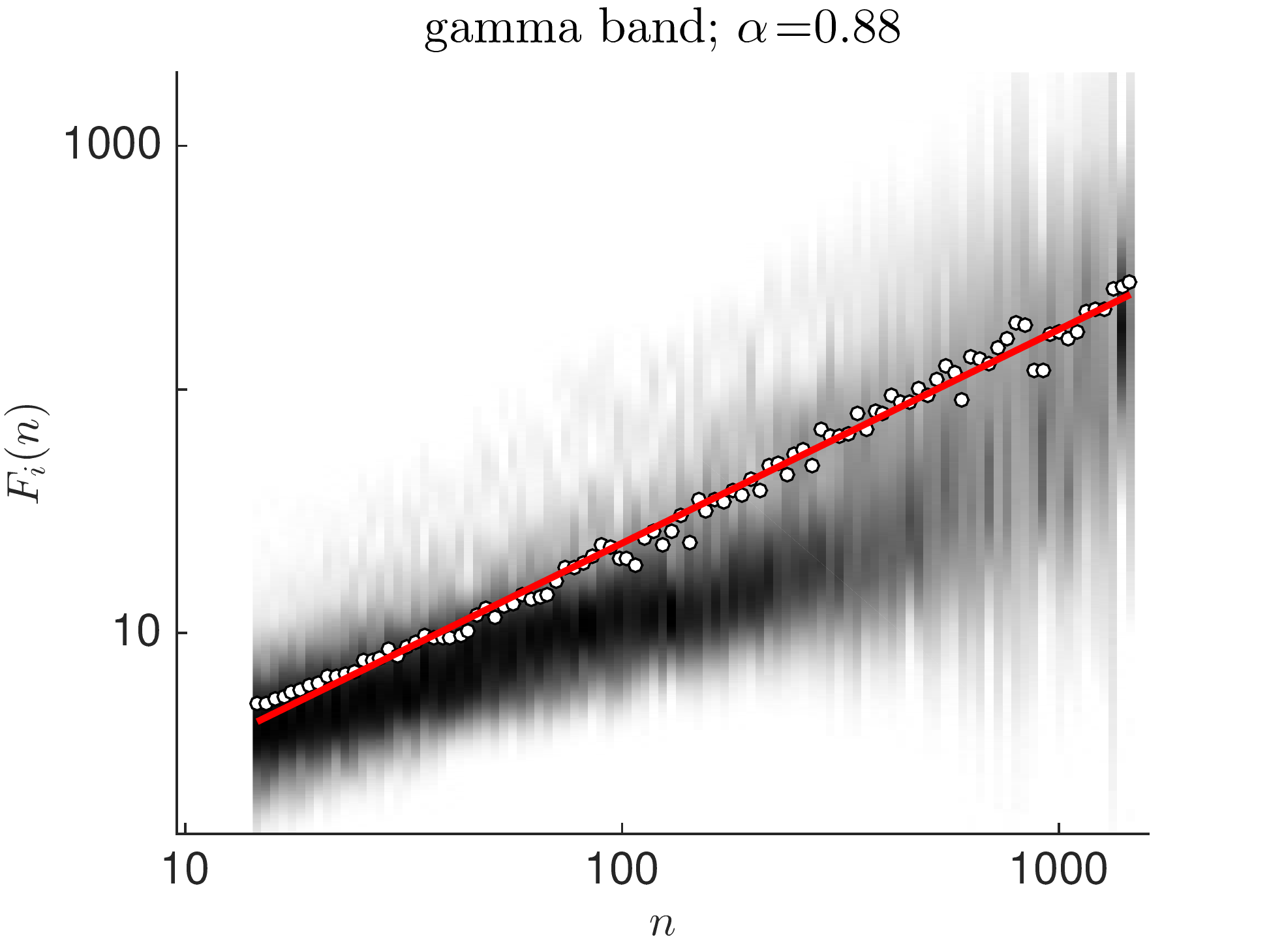}}
\caption{ \small {\bf Fig. \ref{fig:figS5}:} Idem as Fig. \ref{fig:figS1} but for the gamma band. Note in Fig. \ref{subfig:figS5b} the difference in the modes of the $\tilde{p}_n$ distributions indicated by the dark colors and the averaged values $\bar{F}$ in basis of which the fit is performed. This is caused by an asymmetry in the distribution in a similar to Fig. \ref{fig:fig6}. This caused the difference between the $\alpha_{\text{AIC}}/\alpha_{\text{BIC}}$, and the $\bar{\alpha}$ values in Tab. \ref{tab:tabB1}. } \label{fig:figS5}
\end{figure}

The assumption of power-law behavior in these signals was not met in a fairly large number of cases. There the conventional least squares fits differed from our maximum likelihood fits. That is, the conventional $\bar{\alpha}$ values, if at all, are difficult to interpret. We illustrate this in Figs. \ref{fig:figS1}-\ref{fig:figS5}. Fig. \ref{fig:figS5} indicates a possible mechanism responsible the difference in exponent estimates $\bar{\alpha}$ and $\alpha_{\text{AIC}}/\alpha_{\text{BIC}}$ in the alpha and gamma band; the modes of the densities $\tilde{p}_n$ (dark areas) and the averaged values $\bar{F}$ (circles) may differ (fig. \ref{subfig:figS5b}). Since the scaling exponent $\bar{\alpha}$ is estimated on basis of the averaged values, its value is increased compared to the maximum likelihood fit value (not shown in Fig. \ref{subfig:figS5b}). This is similar to the case displayed in Fig. \ref{fig:fig6}, where the discrepancy was caused by the asymmetry in the $\tilde{p}_n$ distributions.\\
\vspace{\fill}
\begin{table}[htb]
\centering
\caption{ \small {\bf Table \ref{tab:tabB1}:} Estimates of power-law exponents for in total 60 MEG signals (six occipital channels and ten subjects). Proportions of the number of signals classified as power law according to AIC$_c$ ($P_{\text{AIC}}$) and BIC ($P_{\text{BIC}}$) agree with the aforementioned observation that BIC tends to penalize more strictly for the number of parameters. Scaling exponents were averaged over all 60 signals ($\tilde{\alpha}$) or over the signals to which a power law was assigned by the BIC ($\alpha_{\text{BIC}}$) or AIC$_c$ ($\alpha_{\text{AIC}}$). The averages over the conventional estimate for these signals classified as a power law are given by $\bar{\alpha}_{\text{AIC}}$ and $\bar{\alpha}_{\text{BIC}}$.} \label{tab:tabB1}
\begin{tabularx}{\linewidth}{X X X X X X}
\toprule
 & delta & theta & alpha & beta & gamma \\
 \hline
$P_{\text{AIC}}$ & 30\% & 63.3\% & 35.0\% & 60.0\% & 56.7\%\\
$\alpha_{\text{AIC}}$ & 0.74 & 0.69 & 0.83 & 0.63 & 0.54 \\
\hline
$P_{\text{BIC}}$ & 63.3\% &   81.7\% &   63.3\% &   68.3\% &   56.7\% \\
$\alpha_{\text{BIC}}$ & 0.72 &    0.68 &    0.82 &    0.63 &    0.54 \\
\hline
$\bar{\alpha}$ & 0.70 &  0.67 &   0.75 &   0.66  &  0.62 \\
$\bar{\alpha}_{\text{AIC}}$ & 0.74 & 0.68 & 0.74 & 0.63 & 0.58 \\
$\bar{\alpha}_{\text{BIC}}$ & 0.71 & 0.67 & 0.74 & 0.63 & 0.58 \\
\bottomrule
\end{tabularx}
\end{table}

\end{appendices}
\end{document}